\documentclass[%
 aip,
 jcp,
 amsmath,amssymb,
reprint,%
]{revtex4-1}

\usepackage{graphicx}
\usepackage{dcolumn}
\usepackage{bm}
\usepackage{amsmath}
\usepackage[hyperindex,breaklinks]{hyperref}
\usepackage[utf8]{inputenc}
\hypersetup{
     colorlinks = true,
     citecolor = blue,
     linkcolor = blue,
     urlcolor = blue
}
\def\eqref#1{\textcolor{blue}{(\ref{#1})}} 

\begin{document}


\title{Stochastic sampling of the isothermal-isobaric ensemble: phase diagram of crystalline solids from molecular dynamics simulation}

\author{Samuel Cajahuaringa}
\email{samuelif@ifi.unicamp.br}
\affiliation{Instituto de F\'{i}sica Gleb Wataghin, Universidade Estadual de Campinas, UNICAMP, 
13083-859 Campinas, S\~{a}o Paulo, Brazil}

\author{Alex Antonelli}
\email{aantone@ifi.unicamp.br}
\affiliation{Instituto de F\'{i}sica Gleb Wataghin and Centre for Computational Engineering \& Sciences, Universidade Estadual de Campinas, UNICAMP, 13083-859 Campinas, S\~{a}o Paulo, Brazil}

\date{\today}

\begin{abstract}
A methodology to sample the isothermal-isobaric ensemble using Langevin dynamics is proposed, which combines novel features of geometric integrators for the equations of motion. By employing the Trotter expansion, the methodology generates a robust, symmetric and accurate numerical algorithm. In order to show that the proposed method correctly samples the phase-space, simulations in the isotropic NPT ensemble were carried out for two analytical examples. Also this method permits to study a solid-solid phase transition, by conducting a fully flexible-cell molecular dynamics simulation. Additionally, we present an efficient method to determine the Gibbs free energy in a wide interval of pressure along an isothermal path, which allows us to determine the transition pressure in a driven by pressure solid-solid phase transition. Our calculations show that the methodology is highly suitable for the study of phase diagram of crystalline solids.
\end{abstract}

\maketitle

\section{Introduction}\label{sec:Introduction}

Molecular Dynamics (MD) is a computer simulation method that has become one of the most important and commonly used approaches to study condensed phase systems.\cite{Allen} The particular advantage of MD is its ability to obtain macroscopic observables of interest in the formalism of statistical mechanics.\cite{Hansen}

MD was originally developed to sample the microcanonical ensemble, i.e., simulations under constant number of particles, volume, and energy (NVE) conditions. However, isolated system conditions are not those in which experiments are performed. In order to reflect experimental conditions, several methods have been developed, in such a way that MD simulations can sample canonical and isothermal-isobaric ensembles.

The most commonly used algorithms for generating the desired ensemble are based on the extended phasespace approach, which is done by coupling the system to additional degrees of freedom representing a thermal bath and/or a barostat, in order to maintain isothermal and/or isobaric conditions.\cite{liouvillenhc,liouvillenhc2} For example, the Nos\'{e}-Hoover\cite{nose,hoover} and Nos\'{e}-Hoover chain\cite{nhc} (NHC) thermostats have been proposed to generate the canonical ensemble. A combination of Nos\'{e}-Hoover thermostats with the Andersen\cite{andersen} or the Parrinello-Rahman\cite{parrinello,MTK} barostats have been proposed to generate the NPT ensemble, with isotropic or anisotropic volume fluctuations, respectively.

In particular for crystallines solids, where the harmonic forces are important, the deterministic thermostat leading to incorrect phase-space sampling\cite{nose,hoover} since it is well known that such thermostat fails to sample correctly a system composed by harmonic oscillators. In order to overcome this difficulty, the so-called massive Nos\'{e}-Hoover chain\cite{nhc} (MNHC) dynamics can be used, which consists in coupling one NHC to every degree of freedom of each particle. One important setback of the MNHC in classical MD simulations is its relative stiffness that requires a shorter time step and the high computational cost to propagate this dynamics, due to the increase of the extended phase-space. Therefore, various deterministic schemes have been developed in order to simulate a harmonic oscillator with one single thermostat.\cite{hoover2,Hoover3,Hoover4,Tapias1,Bravetti} Recently, this difficulty has been overcome successfully for one-dimensional systems by using the logistic thermostat.\cite{Tapias2,Hoover5} This is a promising approach to reduce significantly the extended phase-space to simulate crystalline solids, where harmonic and anharmonic forces are present.

In order to circunved the need to increased the extended phase-space, the other approach involves the use of the stochastic method that mimic the behavior of a thermal bath such as the Langevin dynamics, which is ergodic in the more difficult cases,\cite{Matt} but it is more disruptive of the dynamics.\cite{Bussi2} So it ensures the correct sampling of the canonical ensemble under any circumstances.

Analogously to the extended phase-space approach, it has been proved that the stochastic method is also able to generate the NPT ensemble. In this approach, the Langevin dynamics is considered for both particles and the barostat.\cite{feller,kolb,quigley,jensen,Bussi3,Gao} Several authors have reported equations of motion and integrations schemes to sample the NPT ensemble with isotropic volume fluctuations,\cite{feller,kolb,quigley,jensen,Bussi3} and if possible extended the method to fully flexible cell,\cite{quigley,Gao} i.e., anisotropic volume fluctuations.

In this work, we show how to implement the Langevin dynamics with barostats within the extended phase-space approach, in order to perform isobaric-isothermal simulations. The equations of motion are integrated using stochastic-geometric integrators to control the temperature through the Langevin dynamics, as proposed by Bussi et al.,\cite{bussi} and the deterministic part corresponding to the barostat is integrated using geometric integrators.\cite{MTK,liouvillenhc,liouvillenhc2} The integrator is constructed by using the symmetric Trotter factorization\cite{tuckerman,BookTuckerman}, which is normally employed to decompose the Liouville\cite{tuckerman,BookTuckerman,liouvillenhc,liouvillenhc2,MTK} or the Fokker-Planck\cite{bussi,Bussi3,quigley,Gao} operators which describe the dynamics of an ensemble of realizations.

We show that this kind of NPT integrator can sample the phase-space accurately by simulating two different toy models that can be solved analytically\cite{liouvillenhc,jensen} (with isotropic volume fluctuations), comparing our numerical results with probability distribution functions, and deriving measurable thermodynamic quantities, such as the absolute Gibbs free energy, in order to evidence the ergodicity of the NPT Langevin dynamics. We also remark that the proposed integrator for the fully flexible cell case allows all elements of the cell matrix to vary independently, this avoids the use of approximations in the propagation of the equations of motion, which are needed when a restricted cell matrix is utilized, due to the presence of singularities.\cite{Gao} This feature of our integrator for the flexible-cell case is very important for our purposes to apply the methodology to study solid-solid phase transitions, and represents the main difference between our approach and those used to derive other integrators.\cite{Gao}

Additionally, we also review the methods to estimate free energy differences between two equilibrium states, such as thermodynamic integration\cite{free} and adiabatic switching,\cite{AS} and propose a new methodology to compute the absolute Gibbs free energy difference along an isothermal path. 

To prove the effectivity of the NPT integrator with fully flexible cell, we simulate a two-dimensional system that displays a solid-solid phase transition and determine the transition pressure by the crossing of the Gibbs free energy curves of both phases. 

The article is organized as follows. Sec. \ref{sec:Langevin-NPT} describes in detail the NPT Langevin dynamics. In Sec. \ref{sec:integrating} it is shown how the numerical integrators for the NPT Langevin dynamics are implemented, for the isotropic cell and the fully flexible one, in Subsecs. \ref{sec:npt-iso} and \ref{sec:npt-flex}, respectively. The methodological aspects of the calculation of free energy are described in Sec. \ref{sec:Free-energy}. Sec. \ref{sec:Ilustrative-examples} provides the results that demonstrate the efficiency of sampling the NPT ensemble using the Langevin dynamics. Finally, in Sec. \ref{sec:Conclusions} we summarize the main results of our work.
\section{Equations of motion}\label{sec:Langevin-NPT}
We begin by considering a classical system of $N$ particles in $d$ dimensions with positions 
$\textbf{r}=\{\textbf{r}_{1};\ldots;\textbf{r}_{N}\}$ and momenta $\textbf{p}=\{\textbf{p}_{1};\ldots;\textbf{p}_{N}\}$ in a
general container of volume $V$ described by a cell matrix $\textbf{h}$. 
Let $U(\textbf{r}_{1};\ldots;\textbf{r}_{N})$ denote the interparticle potential and $\textbf{f}_{i}=-\partial U/\partial\textbf{r}_{i}$ denote the forces acting on each particle.

Analogously to the equations of motion of Martyna et al.\cite{MTK} for isotropic volume fluctuations, where two independent thermostats are used to control the kinetics of the particles and the barostat, we also employ two independent Langevin thermostats.\cite{feller,kolb,quigley,jensen} This device is particularly important, since the barostat typically evolves on a slower time scale than that of the particles, and the barostat controls the fluctuations in the internal pressure estimator,
\begin{equation}
P^{int}=\frac{1}{dV}\bigg[\sum_{i=1}^{N}\frac{\textbf{p}_{i}^{2}}{m_{i}}+\sum_{i=1}^{N}\textbf{f}_{i}\cdot\textbf{r}_{i}\bigg]-\frac{\partial U}{\partial V}.
\label{eqn:pressure-iso}
\end{equation}
The term $\partial U/\partial V$ in Eq.~\eqref{eqn:pressure-iso} arises when the potential has an explicit volume dependence, as to permit sampling the isobaric-isothermal ensemble with isotropic volume fluctuations. Then, the equations of motion are:
\begin{align}
&\dot{\textbf{r}}_{i}=\frac{\textbf{p}_{i}}{m_{i}}+\frac{p_{\epsilon}}{W}\textbf{r}_{i}, \nonumber \\
&\dot{\textbf{p}}_{i}=\textbf{f}_{i}-\Big(1+\frac{d}{N_{f}}\Big)\frac{p_{\epsilon}}{W}\textbf{p}_{i}-\gamma\textbf{p}_{i}+\textbf{R}_{i}, \nonumber \\
&\dot{V}=\frac{dV}{W}p_{\epsilon}, \nonumber \\ 
&\dot{p}_{\epsilon}=dV\Big(P^{int}-P^{ext}\Big)+\frac{d}{N_{f}}\sum_{i=1}^{N}\frac{\textbf{p}_{i}^{2}}{m_{i}}-\gamma_{b}p_{\epsilon}+R_{b}, 
\label{eqn:langevin-npt-iso}
\end{align}
where $\gamma$ is the friction coefficient of the particles, $\gamma_{b}$ is the friction coefficient of the barostat, and $N_{f}$ is the number the degrees of freedom of the system. The system's volume $V$ is a dynamical variable of the barostat and the corresponding momentum variable $p_{\epsilon}$ is the strain rate $\dot{\epsilon}$ multiplied by the fictitious mass $W$ of the barostat. $P^{ext}$ is the external pressure. $\textbf{R}_{i} $ and $R_{b}$ are stochastic white noises applied to particles and barostat, respectively, which are usually assumed to be Gaussian distributed. 

The friction coefficients and the stochastic noise are related by the fluctuation-dissipation theorem, which obey the following properties:
\begin{eqnarray}
\langle R_{i\alpha}(t) \rangle &=& 0, \nonumber \\
\langle R_{i\alpha}(t)R_{i\beta}(t') \rangle &=& 2mk_{B}T\gamma\delta(t-t')\delta_{\alpha\beta}, \\
\langle R_{b}(t) \rangle &=& 0, \nonumber \\
\langle R_{b}(t)R_{b}(t') \rangle &=& 2Wk_{B}T\gamma_{b}\delta(t-t').
\end{eqnarray}

We can show that the probability density of the extended phase-space spanned by $\textbf{x}=(\textbf{r},\textbf{p},V,p_{\epsilon})$ satisfies the following Fokker-Planck equation (see Appendix \ref{app:A} for more details):
\begin{equation}
\frac{\partial }{\partial t}\rho(\textbf{r},\textbf{p},V,p_{\epsilon};t)=-\mathcal{L}_{NPT(iso)}\rho(\textbf{r},\textbf{p},V,p_{\epsilon};t),
\end{equation}
whose solution in the stationary regime corresponds to the following ansatz
\begin{equation}
\rho(\textbf{r},\textbf{p},V,p_{\epsilon})\propto e^{-(\mathcal{H}(\textbf{r},\textbf{p})+p^{2}_{\epsilon}/2W+P^{ext}V)/k_{B}T}.
\label{eqn:rho-npt-iso}
\end{equation}

Integrating again the components of the momentum of the barostat $p_{\epsilon}$  gives a constant, yielding the isobaric-isothermal distribution
\begin{equation}
\rho(\textbf{r},\textbf{p},V)\propto e^{-(\mathcal{H}(\textbf{r},\textbf{p})+P^{ext}V)/k_{B}T}.
\label{eqn:rho-iso}
\end{equation}

If anisotropic volume fluctuations are permitted, which change both
size and shape of the box, it is possible to couple the Langevin dynamics to the Parrinello-Rahman barostat.\cite{quigley} In this case the equations of motion take the form:
\begin{align}
&\dot{\textbf{r}}_{i}=\frac{\textbf{p}_{i}}{m_{i}}+\frac{\textbf{p}_{g}}{W_{g}}\textbf{r}_{i},\nonumber \\
&\dot{\textbf{p}}_{i}=\textbf{f}_{i}-\frac{\textbf{p}_{g}}{W_{g}}\textbf{p}_{i}-\frac{1}{N_{f}}\frac{\mathrm{tr}(\textbf{p}_{g})}{W_{g}}\textbf{p}_{i}-\gamma\textbf{p}_{i}+\textbf{R}_{i},\nonumber \\
&\dot{\textbf{h}}=\frac{\textbf{p}_{g}\textbf{h}}{W_{g}}, \\
&\dot{\textbf{p}}_{g}=\mathrm{det}(\textbf{h})\Big(\textbf{P}^{int}-\textbf{I}P^{ext}\Big)+\frac{1}{N_{f}}\sum_{i=1}^{N}\frac{\textbf{p}_{i}^{2}}{m_{i}}\textbf{I}-\gamma_{b}\textbf{p}_{g}+\textbf{R}_{b}, \nonumber
\label{eqn:langevin-npt-flex}
\end{align}
here $\textbf{I}$ is the $d\times d$ identity matrix, where $d$ is the number of dimensions, $\textbf{p}_{g}$ is the $d\times d$ barostat momentum matrix, $W_{g}$ is the barostat mass, $\textbf{h}$ is the $d\times d$ cell matrix, $\textbf{R}_{b}$ is the stochastic force matrix used to control the temperature of the barostat, which is related to the friction coefficient of barostat by the theorem of fluctuation-dissipation
\begin{eqnarray}
\langle R_{b,\alpha\beta}(t) \rangle &=& 0, \nonumber \\
\langle R_{b,\alpha\beta}(t)R_{b,\alpha\beta}(t') \rangle &=& 2W_{g}k_{B}T\gamma_{b}\delta(t-t').
\end{eqnarray}
and $\textbf{P}^{int}$ is the internal pressure tensor, whose components are given by
\begin{align}
\mathrm{P}^{int}_{\alpha,\beta}&=\frac{1}{\mathrm{det}(\textbf{h})}\sum_{i=1}^{N}\bigg[\frac{\mathrm{p}_{i,\alpha}\mathrm{p}_{i,\beta}}{m_{i}}+\mathrm{r}_{i,\alpha}\mathrm{f}_{i,\beta}\bigg] \nonumber \\
&-\frac{1}{\mathrm{det}(\textbf{h})}\sum_{\gamma=1}^{d}\frac{\partial U}{\partial\mathrm{h}_{\alpha,\gamma}}\mathrm{h}_{\gamma,\beta}.
\end{align}

This dynamics is equivalent to the evolution of the following unweighted distribution function (see Appendix \ref{app:B} for more details) of the extended phase-space spanned by $\textbf{x}=(\textbf{r},\textbf{p},\textbf{h},\textbf{p}_{g})$, which corresponds to the following Fokker-Planck equation:
\begin{equation}
\frac{\partial }{\partial t}\rho(\textbf{r},\textbf{p},\textbf{h},\textbf{p}_{g};t)=-\mathcal{L}_{NPT(flex)}\rho(\textbf{r},\textbf{p},\textbf{h},\textbf{p}_{g};t), 
\end{equation}
it can be shown that in the steady regime the following ansatz solves the equation above,
\begin{equation}
\rho(\textbf{r},\textbf{p},\textbf{h},\textbf{p}_{g})\propto e^{-(\mathcal{H}(\textbf{r},\textbf{p})+\frac{\mathrm{tr}(\textbf{p}_{g}^{T}\textbf{p}_{g})}{2W_{g}}+P^{ext}\mathrm{det}(\textbf{h}))/k_{B}T}.
\end{equation}
We need to introduce the following factor $[\mathrm{det}(\textbf{h})]^{1-d}$ in order to obtain the correct distribution function for anisotropic volume fluctuations. This additional factor is essential so that the equations of motion and numerical solver generate this factor as part of the phase-space volume element (see Appendix \ref{app:B} for more details).

The integration over the components of the momentum of the barostat, $\textbf{p}_{g}$, again yields a constant, resulting in the correct isothermal-isobaric partition
function for fully flexible cells
\begin{equation}
\rho(\textbf{r},\textbf{p},\textbf{h})\propto[\mathrm{det}(\textbf{h})]^{1-d}e^{-(\mathcal{H}(\textbf{r},\textbf{p})+P^{ext}\mathrm{det}(\textbf{h}))/k_{B}T}.
\label{eqn:rho-flex}
\end{equation}
\section{Integrating the equations of motion}\label{sec:integrating}
The integration scheme presented in this section allows for the sampling of the isotropic NPT ensemble, which will be verified using two analytical examples, and its generalization to the fully flexible case allows for the sampling of the anisotropic NPT ensemble. The approach we will take is based on the derivation of geometric integrators that was introduced previously in the literature for deterministic\cite{tuckerman,BookTuckerman,liouvillenhc,liouvillenhc2} and stochastic\cite{bussi,quigley} equations of motion.
\subsection{Sampling the NPT(iso) ensemble using Langevin dynamics}\label{sec:npt-iso}
Following Ref.~\onlinecite{quigley}, we decompose the Fokker-Planck operator for the isotropic NPT ensemble (Eq.~\eqref{eqn:fokker-planck-npt-iso}) into the following set of terms:
\begin{equation}
\mathcal{L}_{NPT(iso)}=\mathcal{L}_{1}+\mathcal{L}_{2}+\mathcal{L}_{V}+\mathcal{L}_{p_{\epsilon}}+\mathcal{L}_{\gamma}+\mathcal{L}_{\gamma_{b}}.
\label{eqn:operator-fk-npt-iso}
\end{equation}
Here, the individual contributions in Eq.~\eqref{eqn:operator-fk-npt-iso} are given by
\begin{align}
&\mathcal{L}_{1}=\sum_{i=1}^{N}\bigg[\frac{\textbf{p}_{i}}{m_{i}}+\frac{p_{\epsilon}}{W}\textbf{r}_{i}\bigg]\cdot\frac{\partial }{\partial \textbf{r}_{i}}, \nonumber \\
&\mathcal{L}_{2}=\sum_{i=1}^{N}\bigg[\textbf{f}_{i}-\frac{\alpha p_{\epsilon}}{W}\textbf{p}_{i}\bigg]\cdot\frac{\partial }{\partial \textbf{p}_{i}}, \nonumber \\
&\mathcal{L}_{V}=\frac{dVp_{\epsilon}}{W}\frac{\partial }{\partial V}, \nonumber \\ 
&\mathcal{L}_{p_{\epsilon}}=G_{\epsilon}\frac{\partial }{\partial p_{\epsilon}}, \nonumber \\
&\mathcal{L}_{\gamma}=-\gamma\sum_{i=1}^{N}\frac{\partial}{\partial \textbf{p}_{i}}\cdot\big(\textbf{p}_{i}+m_{i}k_{B}T\frac{\partial}{\partial \textbf{p}_{i}}\big), \nonumber \\
&\mathcal{L}_{\gamma_{b}}=-\gamma_{b}\frac{\partial}{\partial p_{\epsilon}}\big(p_{\epsilon}+Wk_{B}T\frac{\partial}{\partial p_{\epsilon}}\big),
\label{eqn:Ls-npt-iso}
\end{align}
where $\mathcal{L}_{\gamma}$ and $\mathcal{L}_{\gamma_{b}}$ are the stochastic Langevin operators applied to the particles and the barostat, respectively.\cite{bussi} In Eq.~\eqref{eqn:Ls-npt-iso}, $\alpha=1+\frac{d}{N_{f}}$ and
\begin{equation}
G_{\epsilon}=dV\Big(P^{int}-P^{ext}\Big)+\frac{d}{N_{f}}\sum_{i=1}^{N}\frac{\textbf{p}_{i}^{2}}{m_{i}},
\end{equation}
which acts as a force on the barostat variable $p_{\epsilon}$. Given the decomposition in Eq.~\eqref{eqn:operator-fk-npt-iso}, we factorize the single time-step propagator $\exp(-\Delta t\mathcal{L}_{NPT(iso)})$  as follows:
\begin{eqnarray}
e^{-\Delta t\mathcal{L}_{NPT(iso)}}\approx e^{-\Delta t/2\mathcal{L}_{\gamma_{b}}}e^{-\Delta t/2\mathcal{L}_{\gamma}}e^{-\Delta t/2\mathcal{L}_{p_{\epsilon}}}e^{-\Delta t/2\mathcal{L}_{2}}\nonumber \\
\times e^{-\Delta t\mathcal{L}_{V}}e^{-\Delta t\mathcal{L}_{1}}e^{-\Delta t/2\mathcal{L}_{2}}e^{-\Delta t/2\mathcal{L}_{p_{\epsilon}}}e^{-\Delta t/2\mathcal{L}_{\gamma}}e^{-\Delta t/2\mathcal{L}_{\gamma_{b}}},\qquad
\label{eqn:exp-propagator-npt-iso}
\end{eqnarray}
where the operators $\exp(-\Delta t/2\mathcal{L}_{\gamma})$ and $\exp(-\Delta t/2\mathcal{L}_{\gamma_{b}})$ propagate the particles and the barostat momenta, respectively, by $\Delta t/2$ under the Ornstein-Uhlenbeck process,\cite{bussi} 
\begin{equation}
p(t)=e^{-\gamma t}p(0)+\sqrt{(1-e^{-2\gamma t})mk_{B}T}R,
\label{eqn:OU}
\end{equation}
where $R$ is a random number subject to the Gaussian
distribution with vanishing mean and unit variance. The operator $\exp(-\Delta t/2\mathcal{L}_{p_{\epsilon}})$ is a simple translation operator, while $\exp(-\Delta t\mathcal{L}_{V})$ is a scaling operator of the volume,\cite{liouvillenhc,liouvillenhc2} which is carried out once and not twice, as it is implied by the expansion in Eq.~\eqref{eqn:exp-propagator-npt-iso}. This is because $\mathcal{L}_{1}$ and $\mathcal{L}_{V}$ are commuting operators, since they act on decoupled variables, and, therefore, the two half-step integrations of $\mathcal{L}_{V}$ on either sides of $\mathcal{L}_{1}$ can be combined into one step integration. The action of the operator $\exp(\mathcal{L}_{1}\Delta t)$ can be determined by solving the differential equation:
\begin{equation}
\dot{\textbf{r}}_{i}=\textbf{v}_{i}+v_{\epsilon}\textbf{r}_{i}, 
\end{equation}
for constant $\textbf{v}_{i}=\textbf{p}_{i}/m_{i}$ and $v_{\epsilon}=p_{\epsilon}/W$, for an arbitrary initial condition $\textbf{r}_{i}(0)$, and evaluating the solution at $t=\Delta t$, it yields the evolution,
\begin{equation}
\textbf{r}_{i}(\Delta t)=\textbf{r}_{i}(0) e^{v_{\epsilon}\Delta t} + \Delta t \textbf{v}_{i}e^{v_{\epsilon}\Delta t/2}\frac{\sinh(v_{\epsilon}\Delta t/2)}{v_{\epsilon}\Delta t/2}.
\label{eqn:x-iso}
\end{equation}

Similarly, the action of operator $\exp(\mathcal{L}_{2}\Delta t/2)$ can be determined by solving the differential equation
\begin{equation}
\dot{\textbf{v}}_{i}=\frac{\textbf{f}_{i}}{m_{i}}-\alpha v_{\epsilon}\textbf{v}_{i},
\end{equation}
for constant $\textbf{f}_{i}$, an arbitrary initial condition $\textbf{v}_{i}(0)$, and evaluating
the solution at $t=\Delta t/2$, it yields the finite-difference expression
\begin{equation}
\textbf{v}_{i}(\Delta t/2)=\textbf{v}_{i}(0) e^{-\alpha v_{\epsilon}\Delta t/2} + \frac{\Delta t}{2m_{i}} \textbf{f}_{i}e^{-\alpha v_{\epsilon}\Delta t/4}\frac{\sinh(\alpha v_{\epsilon}\Delta t/4)}{\alpha v_{\epsilon}\Delta t/4}.
\end{equation}

In practice, the factor $\sinh(x)/x$ should be evaluated by a power series for small $x$ to avoid numerical instabilities, which can be accomplished using the suitable expansion for $\frac{\sinh(x)}{x}\approx\displaystyle\sum_{i=0}^{5}a_{2i}x^{2i}$ where $a_{0}=1$, $a_{2}=1/6$, $a_{4}=1/120$, $a_{6}=1/5040$, $a_{8}=1/362880$, $a_{10}=1/39916800$. These equations together with the Langevin propagator operators completely define an integrator for the isobaric-isothermal ensemble with isotropic volume fluctuations.
\subsection{Sampling the NPT(flex) ensemble using Langevin dynamics}\label{sec:npt-flex}
The integrator for the anisotropic case employs the same basic factorization scheme as in the isotropic case. First, we decompose the Fokker-Planck operator Eq.~\eqref{eqn:fokker-planck-npt-flex} as
\begin{equation}
\mathcal{L}_{NPT(flex)}=\mathcal{L}_{1}+\mathcal{L}_{2}+\mathcal{L}_{\mathbf{h}}
+\mathcal{L}_{\mathbf{p}_{g}}+\mathcal{L}_{\gamma}+\mathcal{L}_{\gamma_{b}},
\label{eqn:operator-fk-npt-flex}
\end{equation}
where,
\begin{align}
&\mathcal{L}_{1}=\sum_{i=1}^{N}\bigg[\frac{\textbf{p}_{i}}{m_{i}}+\frac{\textbf{p}_{g}}{W_{g}}\textbf{r}_{i}\bigg]\cdot\frac{\partial}{\partial \textbf{r}_{i}}, \nonumber \\
&\mathcal{L}_{2}=\sum_{i=1}^{N}\bigg[\textbf{f}_{i}-\frac{\textbf{p}_{g}}{W_{g}}\textbf{p}_{i}-\frac{1}{N_{f}}\frac{\mathrm{Tr}(\textbf{p}_{g})}{W_{g}}\textbf{p}_{i}\bigg]\cdot\frac{\partial}{\partial \textbf{p}_{i}}, \nonumber \\
&\mathcal{L}_{\mathbf{h}}=\bigg(\frac{\textbf{p}_{g}\textbf{h}}{W_{g}}\bigg)\cdot\frac{\partial}{\partial\textbf{h}}, \nonumber \\
&\mathcal{L}_{\mathbf{p}_{g}}=\textbf{G}_{g}\cdot\frac{\partial}{\partial\textbf{p}_{g}}, \nonumber\\
&\mathcal{L}_{\gamma}=-\gamma\sum_{i=1}^{N}\frac{\partial}{\partial \textbf{p}_{i}}\cdot\big(\textbf{p}_{i}+m_{i}k_{B}T\frac{\partial}{\partial \textbf{p}_{i}}\big), \nonumber \\
&\mathcal{L}_{\gamma_{b}}=-\gamma_{b}\frac{\partial}{\partial \textbf{p}_{g}}\cdot\big(\textbf{p}_{g}+W_{g}k_{B}T\frac{\partial}{\partial \textbf{p}_{g}}\big),
\label{Ls-npt-flex}
\end{align}
and
\begin{equation}
\textbf{G}_{g}=\mathrm{det}(\textbf{h})\Big(\textbf{P}^{int}-\textbf{I}P^{ext}\Big)+\frac{1}{N_{f}}\sum_{i=1}^{N}\frac{\textbf{p}_{i}^{2}}{m_{i}}\textbf{I}.
\end{equation}
The $\mathcal{L}_{\gamma}$ and $\mathcal{L}_{\gamma_{b}}$ are the usual stochastic Langevin operators for the particles and for the barostat in the full flexible cell case, respectively. The propagator is factorized exactly as in Eq.~\eqref{eqn:exp-propagator-npt-iso} with the operators $\mathcal{L}_{V}$ and $\mathcal{L}_{p_{\epsilon}}$ replaced by $\mathcal{L}_{\mathbf{h}}$ and $\mathcal{L}_{\mathbf{p}_{g}}$, respectively. It is important to note that $\mathcal{L}_{1}$ and $\mathcal{L}_{\mathbf{h}}$ commute. The overall factorized propagator then becomes,
\begin{eqnarray}
e^{-\Delta t\mathcal{L}_{NPT(flex)}}\approx e^{-\Delta t/2\mathcal{L}_{\gamma_{b}}}e^{-\Delta t/2\mathcal{L}_{\gamma}}e^{-\Delta t/2\mathcal{L}_{\mathbf{p}_{g}}}e^{-\Delta t/2\mathcal{L}_{2}}\nonumber\\
\times e^{-\Delta t\mathcal{L}_{\mathbf{h}}}e^{-\Delta t\mathcal{L}_{1}}e^{-\Delta t/2\mathcal{L}_{2}}e^{-\Delta t/2\mathcal{L}_{\mathbf{p}_{g}}}e^{-\Delta t/2\mathcal{L}_{\gamma}}e^{-\Delta t/2\mathcal{L}_{\gamma_{b}}}.\qquad
\label{eqn:exp-propagator-npt-flex}
\end{eqnarray}

The evaluation of the action of the operators, $\exp(\mathcal{L}_{1}\Delta t)$ and $\exp(\mathcal{L}_{2}\Delta t/2)$, requires the solution of the following matrix-vector differential equations:
\begin{eqnarray}
\dot{\textbf{r}}_{i}&=&\textbf{v}_{i}+\textbf{v}_{g}\textbf{r}_{i} \label{eqn:L1-flex}, \\
\dot{\textbf{v}}_{i}&=&\frac{\textbf{f}_{i}}{m_{i}}-\textbf{v}_{g}\textbf{v}_{i}-b\mathrm{Tr}(\textbf{v}_{g})\textbf{v}_{i}, 
\label{eqn:L2-flex}
\end{eqnarray}
for constants $\textbf{f}_{i}$ and $\textbf{v}_{g}$, where for convenience $\textbf{v}_{g}=\frac{\textbf{p}_{g}}{W_{g}}$ and $b=1/\mathrm{N_{f}}$. In order to solve Eq.~\eqref{eqn:L1-flex}, we introduce the transformation:
\begin{eqnarray}
\mathbf{x}_{i}&=&\mathbf{O}^{\mathrm{T}}\mathbf{r}_{i}, \nonumber \\
\mathbf{u}_{i}&=&\mathbf{O}^{\mathrm{T}}\mathbf{v}_{i},
\end{eqnarray}
where $\mathbf{O}^{\mathrm{T}}$ is an orthogonal matrix, which satisfies $\mathbf{O}^{\mathrm{T}}\mathbf{O}=\mathbf{I}$. The application of this transformation to Eq.~\eqref{eqn:L1-flex} yields
\begin{eqnarray}
\mathbf{O}^{\mathrm{T}}\dot{\textbf{r}}_{i}&=&\mathbf{O}^{\mathrm{T}}\textbf{v}_{i}+ \mathbf{O}^{\mathrm{T}}\textbf{v}_{g}\textbf{r}_{i}, \nonumber \\
\dot{\mathbf{x}}_{i}&=&\mathbf{u}_{i}+\mathbf{O}^{\mathrm{T}}\textbf{v}_{g}\mathbf{O}\mathbf{x}_{i}. \label{eqn:xi-flex}
\end{eqnarray}

Now, by considering a symmetric pressure tensor, it implies that $\textbf{v}_{g}$ is also symmetric. Therefore, it is possible to choose $\mathbf{O}$ to be the orthogonal matrix that diagonalizes $\textbf{v}_{g}$  according to
\begin{equation}
\textbf{v}^{d}_{g}=\mathbf{O}^{\mathrm{T}}\textbf{v}_{g}\mathbf{O},
\end{equation}
where $\textbf{v}^{d}_{g}$ is a diagonal matrix with the eigenvalues of $\textbf{v}_{g}$ in its diagonal. The columns of $\mathbf{O}$ are just the orthonormalized eigenvectors of $\textbf{v}_{g}$. Let $\lambda_{\alpha}$, $\alpha=1,2,3$, be the eigenvalues of $\textbf{v}_{g}$. Since $\textbf{v}_{g}$ is symmetric, its eigenvalues are real. In this representation, the three components of $\textbf{x}_{i}$ are uncoupled in Eq.~\eqref{eqn:xi-flex} and can be solved independently using Eq.~\eqref{eqn:x-iso}. Transforming back to $\textbf{r}_{i}$, we find
\begin{equation}
\textbf{r}_{i}(\Delta t)=\mathbf{O}\mathbf{D}_{1}\mathbf{O}^{\mathrm{T}}\textbf{r}_{i}(0)+ \Delta t\mathbf{O}\mathbf{D}_{2}\mathbf{O}^{\mathrm{T}}\textbf{v}_{i}(0),
\end{equation}
where the matrices $\mathbf{D}_{1}$ and $\mathbf{D}_{2}$ have the elements
\begin{eqnarray}
D_{1,\alpha\beta}&=&e^{\lambda_{\alpha}\Delta t}\delta_{\alpha\beta}, \nonumber \\
D_{2,\alpha\beta}&=&e^{\lambda_{\alpha}\Delta t/2}\frac{\sinh(\lambda_{\alpha}\Delta t/2)}{\lambda_{\alpha}\Delta t/2}\delta_{\alpha\beta}.
\end{eqnarray}

In a similar manner, Eq.~\eqref{eqn:L2-flex} can be solved for $\textbf{v}_{i}$ and the solution evaluated at $t=\Delta t/2$ gives the result,
\begin{equation}
\textbf{v}_{i}(\Delta t/2)=\mathbf{O}\mathbf{\Delta}_{1}\mathbf{O}^{\mathrm{T}}\textbf{v}_{i}(0)+\frac{\Delta t}{2m_{i}}, \mathbf{O}\mathbf{\Delta}_{2}\mathbf{O}^{\mathrm{T}}\textbf{f}_{i}(0),
\end{equation}
where the matrices $\mathbf{\Delta}_{1}$ and $\mathbf{\Delta}_{2}$ have the elements:
\begin{equation}
\Delta_{1,\alpha\beta} = e^{-(\lambda_{\alpha}+b\mathrm{Tr}(\textbf{v}_{g}))\Delta t/2}\delta_{\alpha\beta},  \\
\end{equation}
\begin{equation}
\Delta_{2,\alpha\beta} = e^{-(\lambda_{\alpha}+b\mathrm{Tr}(\textbf{v}_{g}))\Delta t/4}\frac{\sinh((\lambda_{\alpha}+b\mathrm{Tr}(\textbf{v}_{g}))\Delta t/4)}{(\lambda_{\alpha}+b\mathrm{Tr}(\textbf{v}_{g}))\Delta t/4}\delta_{\alpha\beta}.
\end{equation}

The propagators $\exp(\mathcal{L}_{\mathbf{h}}\Delta t)$ and $\exp(\mathcal{L}_{\mathbf{p}_{g}}\Delta t/2)$ can be coded via scaling and direct translation technique, respectively. The operators $\exp(-\Delta t/2\mathcal{L}_{\gamma})$ and $\exp(-\Delta t/2\mathcal{L}_{\gamma_{g}}\Delta t/2)$ propagate the particles and the barostat of full flexible cell (Eq.~\eqref{eqn:OU}), respectively.

In this way, the complete integrator is defined for the anisotropic NPT Langevin dynamics. A technical comment is in order at this point, in order to eliminated cell rotations,\cite{BookTuckerman} it was considered a symmetric pressure tensor, for this reason we only need to thermostat and propagate the diagonal component and the upper or lower non-diagonal elements, to ensure the symmetric behavior of $\textbf{v}_{g}$.
\section{Free Energy Methods}\label{sec:Free-energy}
Several fundamental processes in nature are determined by free energy, whose calculation requires the correct sampling of the phase space. According to our previous discussion, the NPT Langevin dynamics permits the sampling of the isothermal-isobaric ensemble, therefore, the calculation of free energies of various systems and processes is a way to assess the correctness of the sampling provided by the NPT Langevin dynamics. We now briefly review the basics of the methodology we use to compute free energies.\cite{free} Since free energy is a thermal quantity that cannot be expressed in terms of an ensemble average, it cannot be computed directly using Monte Carlo (MC) or Molecular Dynamics (MD) sampling methods. As a result, free energies are usually determined using indirect strategies, in which free energy differences between two systems can be computed by evaluating the work associated with a reversible process along a path connecting a physical system of interest to a reference system. Usually, such path is constructed using a composite Hamiltonian $\mathcal{H}(\lambda)$ coupling the two systems through a parameter $\lambda$. 
Let $F(\lambda)$ be the free energy of the system described by the Hamiltonian $\mathcal{H}(\lambda)$, i.e.
\begin{equation}
F(\lambda)=-k_{B}T\ln\bigg[\frac{1}{(2\pi\hbar)^{3N}}\int_{V}d\textbf{r}d\textbf{p}e^{-\mathcal{H}(\textbf{r},\textbf{p};\lambda)/k_{B}T}\bigg].\label{eqn:dF}
\end{equation}

Let us consider a typical functional form of this coupling given by the Hamiltonian
\begin{equation}
\mathcal{H}(\lambda)=(1-\lambda)\mathcal{H}_{sys}+\lambda\mathcal{H}_{ref},
\end{equation}
where $\mathcal{H}_{sys}$ and $\mathcal{H}_{ref}$ represent the Hamiltonian of the system of interest and that of the reference system, respectively. By taking the derivative of Eq.~\eqref{eqn:dF} with respect to the parameter $\lambda$ and allowing a continuous switching between $\mathcal{H}_{sys}$ and $\mathcal{H}_{ref}$ by varying the parameter $\lambda$ between 0 and 1, the free-energy difference between the two systems determined by the thermodynamic integration (TI) method\cite{free} is
\begin{equation}
F_{sys}-F_{ref}=W_{rev}=\int_{0}^{1}d\lambda\bigg\langle \frac{\partial\mathcal{H}}{\partial\lambda}\bigg\rangle_{\lambda},
\end{equation}
where the brackets indicate the equilibrium average in a statistical ensemble. This integration gives the reversible work $W_{rev}$ done by the generalized force $\partial\mathcal{H}/\partial\lambda$. Since it involves equilibrium averages of the system at all times, it reflects a reversible process. While the TI method is exact in principle, it requires several equilibrium simulations (at least one for each value of $\lambda$) to obtain accurate results.

The adiabatic switching (AS) method\cite{AS,AS2} is derived from the TI method, but is often more efficient for computing free-energy differences, which are estimated from the work integral along a single non-equilibrium simulation during which the value of $\lambda(t)$ changes dynamically, in such a way that, at the beginning of the simulation $\lambda(0)=0$ and at the end $\lambda(t_{sim})=1$ ($t_{sim}$ is the total duration of the switching process). In this case, the reversible work $W_{rev}$ is estimated in terms of the irreversible work $W_{irr}$ estimator
\begin{equation}
W_{irr}=\int_{0}^{t_{sim}}dt'\frac{d\lambda}{dt}\bigg|_{t'}\frac{\partial\mathcal{H(\lambda)}}{\partial\lambda}\bigg|_{\lambda(t')}.
\end{equation}

Given the intrinsic irreversible nature of the process, this dynamic estimator is biased, subject to both statistical and systematic errors. The statistical errors are due to the fact that $W_{irr}$ is a stochastic quantity that depends on the initial conditions, while the systematic error is associated with the dissipative entropy production characteristic of irreversible processes. In this case, due to dissipation, $W_{irr}\geq W_{rev}$. Dissipation effects, however, can be controlled by the simulation time and how the coupling parameter varies with time.\cite{AS2} In this case, within the linear response approximation, the systematic error is independent of the switching process direction, i.e., the entropy production is equal for the forward and backward, processes.\cite{AS3}

In this work, we describe a formulation of the AS method that allows the evaluation of Gibbs free energies at a given temperature over a wide pressure interval from a single simulation. We show that the method is more suitable for the study of solid-solid phase transitions, since this kind of transition is driven by a variation in the external pressure applied on the system. Let us consider a system at two different external pressures, but at the same temperature. It is possible to compute the Gibbs free energy difference between these two different external pressures by evaluating the work associated with a reversible process along a path connecting the physical system at the reference external pressure $P_{0}$ to the system at the external pressure of interest $P$. The corresponding Gibbs free energies are given by
\begin{align}
G_{ref}(N,P_{0},T)&=-k_{B}T\ln\bigg[\frac{1}{(2\pi\hbar)^{3N}}\int_{0}^{\infty}dVe^{-P_{0}V/k_{B}T}\nonumber \\
&\times\int_{V}d\textbf{r}d\textbf{p}e^{-\mathcal{H}(\textbf{r},\textbf{p})/k_{B}T}\bigg],
\end{align}
is the Gibbs free energy at the reference pressure, and
\begin{align}
G_{ref}(N,P,T)&=-k_{B}T\ln\bigg[\frac{1}{(2\pi\hbar)^{3N}}\int_{0}^{\infty}dVe^{-PV/k_{B}T}\nonumber \\
&\times\int_{V}d\textbf{r}d\textbf{p}e^{-\mathcal{H}(\textbf{r},\textbf{p})/k_{B}T}\bigg],
\label{eqn:dG}
\end{align}
is the Gibbs free energy at the pressure of interest. 

The derivative of Eq.~\eqref{eqn:dG} with respect to the external pressure P results in the equilibrium average of the volume in the isothermal-isobaric ensemble. By integrating this result with respect to the external pressure, between the $P_{0}$ and $P$ at constant temperature, one obtains the Gibbs free-energy difference along the isothermal path (TI method), which is given by
\begin{equation}
G_{sys}(N,P,T)-G_{ref}(N,P_{0},T)=W_{mech}=\int_{P_{0}}^{P}dP^{ext}\langle V\rangle,
\label{eqn:Wmech}
\end{equation}
where the brackets in Eq.~\eqref{eqn:Wmech} indicate the equilibrium average of the volume in the isothermal-isobaric ensemble. This integration gives the mechanical work $W_{mech}$ done by the applied external pressure, this relationship is also valid for the fully flexible cell case.

We can estimate the $W_{mech}$ integral along a single non-equilibrium simulation during which the value of external pressure $P^{ext}(t)$ changes dynamically, in such a way that, at the beginning of the simulation $P^{ext}(0)=P_{0}$ and at the end $P^{ext}(t_{sim})=P$. In this case the mechanical work $W_{mech}$ is estimated in terms of the irreversible work $W_{irr}$ determined by the AS method
\begin{equation}
W_{irr}=\int_{0}^{t_{sim}}dt'\frac{dP^{ext}(t)}{dt}\bigg|_{t'}V(t').
\end{equation}

In this switching process not only the initial and final points on the trajectory correspond to physical systems, as it is usual in AS simulations, but the information gathered at the intermediate states of the path also has physical meaning. As a consequence, one obtains the Gibbs free energy of a system in a wide interval of pressure using a one single simulation, provided that the Gibbs free energy of the system of interest in a reference state is known.
\section{Illustrative examples of the NPT Langevin dynamics}\label{sec:Ilustrative-examples}
As a test of the accuracy of the methodology, we have performed simulations with three different models. Initially, we have applied the methodology to two analytical models (toy-models) in the isotropic isobaric-isothermal ensemble. The first system is one particle in an one-dimensional potential\cite{liouvillenhc} and the second one is a particular non-trivial one-dimensional particle model for which one can analytically derive measurable thermodynamic quantities.\cite{jensen} This last example is used as a test model to compute the Gibbs free energy along an isothermal path. These models, therefore, serve as a strict benchmark for the statistical accuracy of a numerical test simulation. The third model is a two-dimensional system that exhibits a solid-solid phase transition,\cite{sspt} which is used to illustrate the anisotropic isobaric-isothermal ensemble and to determine the pressure of the solid-solid phase transition at a given temperature. The transition pressure is obtained by the crossing of the Gibbs free energy curves for each one of the phases as functions of pressure, which are obtained from the new methodology to compute the Gibbs free energy along isothermal paths.%
\subsection{Toy-model I}
In order to illustrate the Langevin NPT integrator's potential and evaluate its performance, we consider the simple example of a single particle with coordinate $q$
\begin{figure}[bp]
    \centering
    \begin{minipage}[b]{1.0\linewidth}
        \includegraphics[scale=0.40,bb=18 75 564 499,clip]{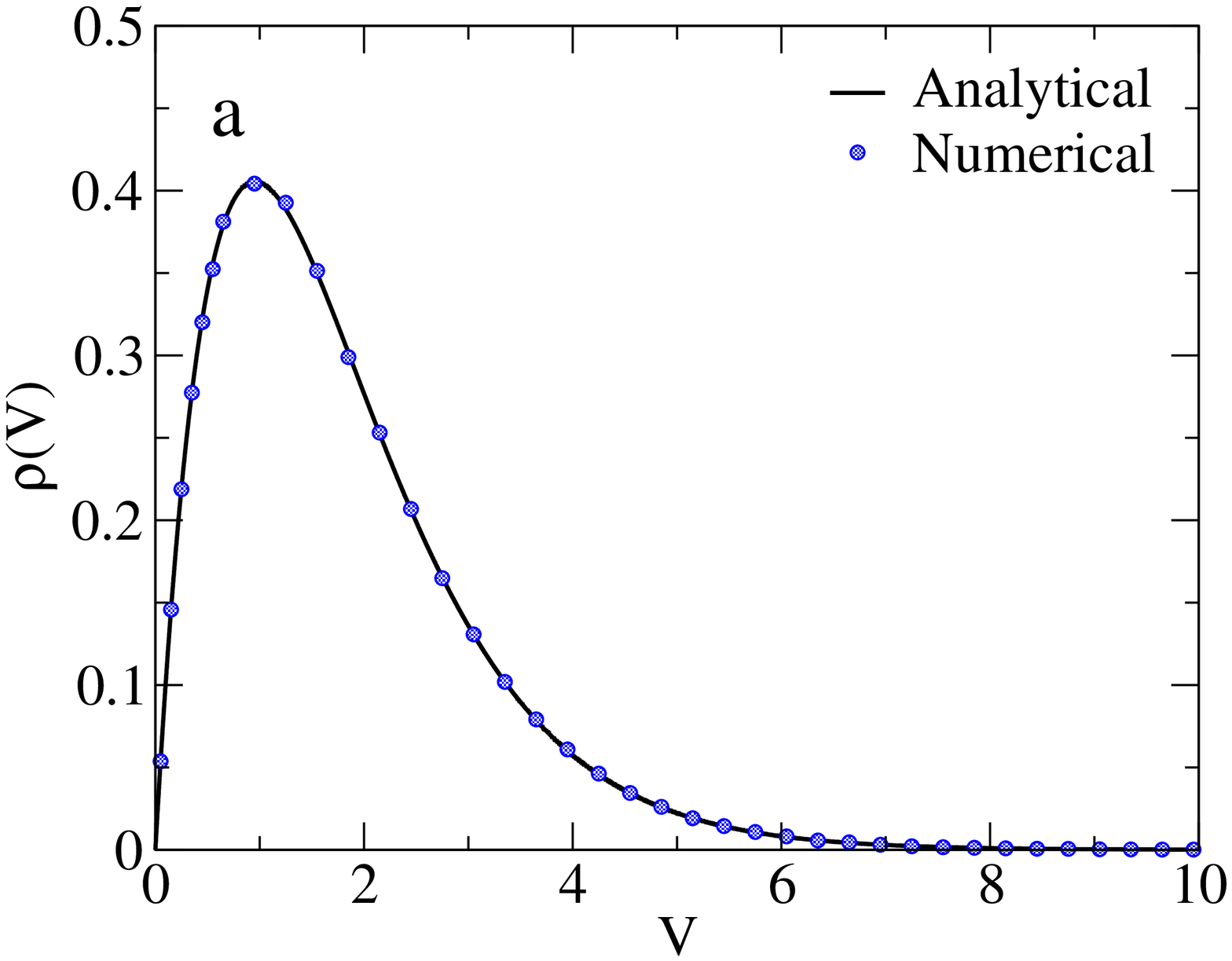}
    \end{minipage}
    \begin{minipage}[b]{1.0\linewidth}
        \includegraphics[scale=0.40,bb=6 75 564 499,clip]{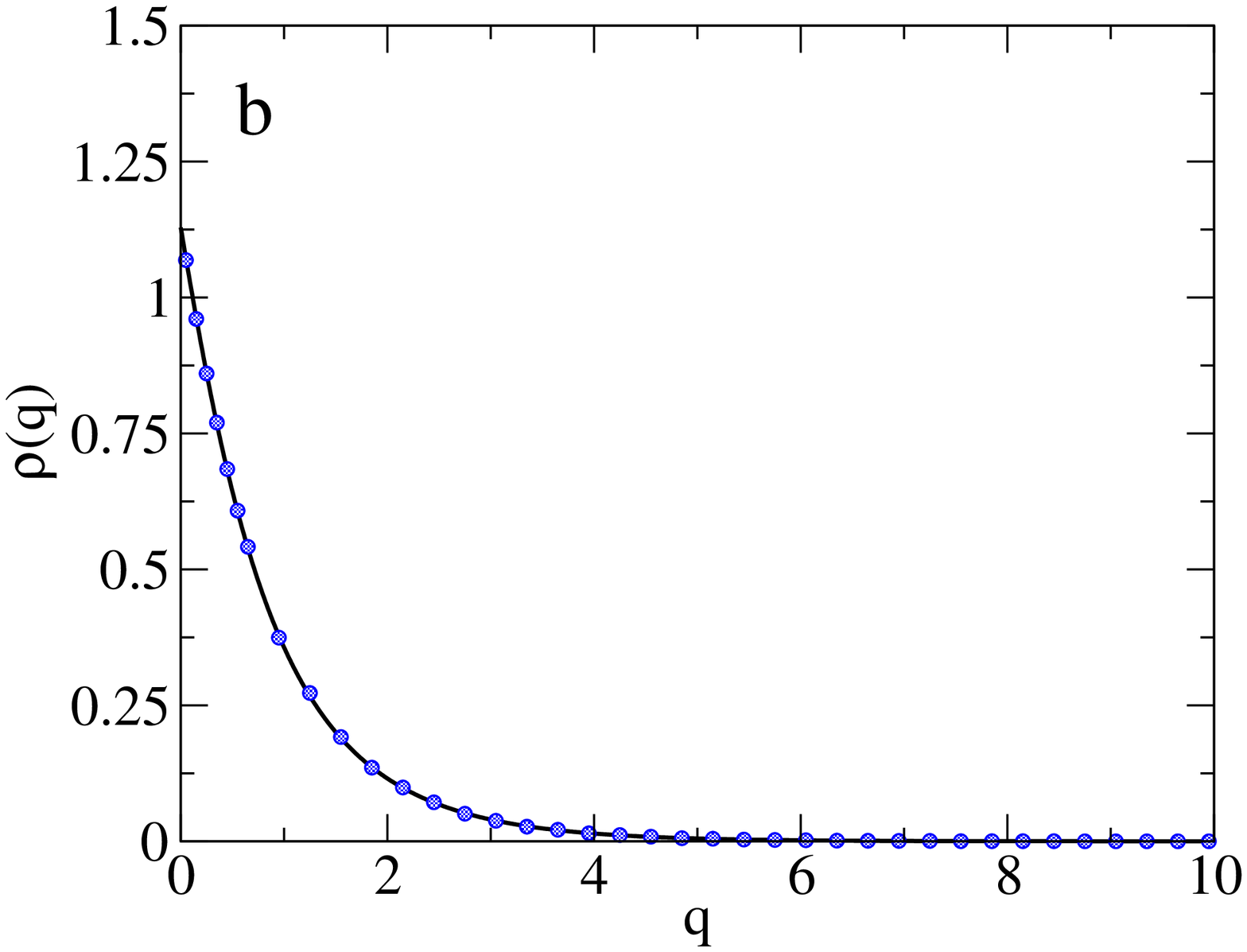}
    \end{minipage}
    \caption{\label{fig:npt-lh} (\textbf{a}) Volume and (\textbf{b}) position distribution functions from the numerical simulation (blue points) and the analytical solutions (solid black line).}
\end{figure}
and momentum $p$ moving in a one-dimensional periodic potential\cite{liouvillenhc} of the form:
\begin{equation}
U(q,V)=\frac{m\omega^{2}V^{2}}{4\pi^{2}}\bigg[1-\cos\Big(\frac{2\pi q}{V}\Big)\bigg]
\label{eqn:example1-potential}.
\end{equation}

Here, $V$ is the volume of the one-dimensional box. The NPT equations of motion for this problem are integrated using a time step of $\Delta t=0.01$. Other parameters are chosen to be $T=1$, $P=1$, $m=1$, $\omega=1$, $k_{B}=1$, $\gamma=0.5$, $\gamma_{b}=0.3333$ and $W=18$. Fig.~\ref{fig:npt-lh} shows the position and volume distributions generated from the simulation together with the corresponding curves obtained analytically. It can be seen from the Fig.~\ref{fig:npt-lh} that these numerical distributions are in excellent agreement with the analytical ones.
\subsection{Toy model-II}
We now present the performance of the integration scheme in the case of another one-dimensional model,\cite{jensen} for which the Gibbs free energy can be calculated analytically. The system consists of $N$ identical particles within a box of length $L$ with periodic boundary conditions, located in the following order $\lbrace{1},x_{1},x_{2},\ldots,x_{N}\rbrace$. Each particle interacts with its two neighbors through the following normalized potential $u(r)$ (in thermal power units $E_{0}=k_{B}T$)
\begin{equation}
u(r)=\frac{\epsilon}{r}+\frac{1}{2}ln(r),
\end{equation}
where $r$ is the normalized pair distance and the connection with the physical potential $U$ is given by $U(r_{0}r)=E_{0}u(r)$ .

In the NPT ensemble the partition function of the system is
\begin{equation}
\mathcal{Z}_{NPT}=\int\mathrm{d}L\prod_{i=1}^{N}\int\mathrm{d}x_{i}\exp\bigg[-\sum_{i=1}^{N}u(x_{i+1}-x_{i})-PL\bigg],
\label{eqn:ZNPT}
\end{equation}
where $P$ in this case denotes the one-dimensional pressure, which is expressed in units of $E_{0}/r_{0}$.

The partition function given by Eq.~\eqref{eqn:ZNPT} can be integrated, resulting in
\begin{equation}
\mathcal{Z}_{NPT}=\bigg[\sqrt{\frac{\pi}{P}}e^{-2\sqrt{P\epsilon}}\bigg]^{N}.
\end{equation}

The Gibbs free energy (in thermal power units) is given by 
\begin{equation}
G=-N\ln\bigg[\sqrt{\frac{\pi}{P}}e^{-2\sqrt{P\epsilon}}\bigg]\label{eqn:Gibbs-jensen}.
\end{equation}

The average length per particle is
\begin{equation}
\langle l \rangle\equiv\frac{\langle L \rangle}{N}=\frac{1}{N}\frac{\partial G}{\partial P}=\frac{1}{2P}+\sqrt{\frac{\epsilon}{P}},
\label{eqn:l-mean}
\end{equation}
and the variance of the normalized length distribution is given by
\begin{equation}
\sigma_{l}^{2}\equiv\frac{\langle (L-\langle L \rangle)^{2}}{N}=-\frac{\partial\langle l\rangle}{\partial P}=\frac{1}{2P^{2}}+\frac{\sqrt{\epsilon}}{2}P^{-3/2}.
\label{eqn:sigma}
\end{equation}

We performed simulations of systems containing $N=1000$ particles for 4 million time steps, using the following parameters: $\epsilon=10$, $k_{B}T=1$, $m=1$, $\gamma=1.0$, $\gamma_{b}=0.01$ for different pressures and time steps. In addition, the friction coefficients of the thermostat $\gamma, \gamma_{b}$ and the barostat mass $W$ were determined by the recommended formulas,\cite{liouvillenhc,martyna,Bussi2}
\begin{eqnarray}
\gamma&=&\frac{1}{2\tau_{p}}, \nonumber \\
\gamma_{b}&=&\frac{1}{10\tau_{b}}, \nonumber \\
W&=&(dN+d)k_{B}T\tau^{2}_{b},
\end{eqnarray}
where $d$ is the dimension of the system, $\tau_{p}$ and $\tau_{b}$ are characteristic time scales of the particles and barostat, respectively. Typically, $\tau_{p}$ and $\tau_{b}$ can be chosen to be directly proportional to the time step $\Delta t$.\cite{martyna} 
\begin{figure}[tbp]
    \centering
        \includegraphics[scale=0.34,bb=11 37 718 548,clip]{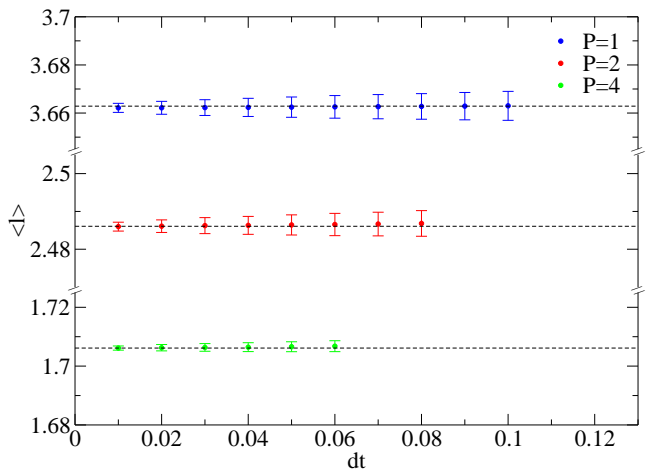}      
    \caption{\label{fig:npt-jensen} Results of the average length per particle at different pressures as function of the time step, the dotted line corresponds to theoretical values.}
\end{figure}

Fig.~\ref{fig:npt-jensen} shows the average length per particle at three different pressures as a function of the time step. The perfect agreement between the simulated and analytical results, shows the robustness of the methodology we developed to sample the isobaric-isothermal ensemble in accordance with the predicted results by the Eq.~\eqref{eqn:l-mean}. 

To illustrate the capabilities of our methodology, we also calculate the Gibbs free energy difference of this system along an isothermal path for a wide interval of pressure. Using as a reference the Gibbs free energy of the system at pressure $P=1$. The equations of motion of the isothermal-isobaric ensemble were integrated using a time step of $\Delta t=0.01$. 
\begin{figure}[tbp]
    \centering
        \includegraphics[scale=0.40,bb=40 71 581 523,clip]{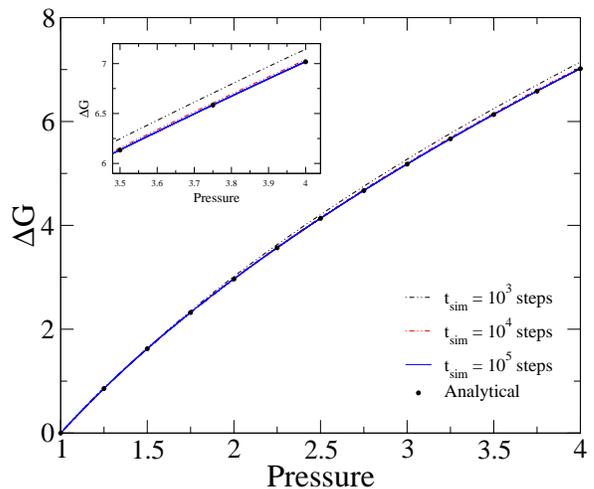}      
    \caption{\label{fig:Gibbsjensen} Convergence of the AS-MD method to compute Gibbs free energy difference per atom along the isothermal path, obtained using three runs with $10^{3}$, $10^{4}$ and $10^{5}$ time steps, respectively. Circles are the analytical values of the Gibbs free energy difference per atom (Eq.~\eqref{eqn:Gibbs-jensen}). (\textbf{Inset}) Gibbs free energy difference per particle at high pressure. The Gibbs free energy at $P=1$ is taken as a reference.}
\end{figure}

The Fig.~\ref{fig:Gibbsjensen} displays the convergence of the Gibbs free energy difference per particle for three values of simulation time $t_{sim}$. 

In the simulations, pressure is scaled according to a linear interpolation within the pressure interval spanning between 1.0 and 4.0, during a switching time equal to the simulation time $t_{sim}$. Fig.~\ref{fig:Gibbsjensen} shows the results of three different AS-MD simulations along the isothermal path, characterized by switching times $t_{sim}$ of $10^{3}$, $10^{4}$ and $10^{5}$ steps, respectively. Each curve shows the Gibbs free energy difference per particle (in units of $k_{B}T$) as a function of pressure. The circles represent analytical values of $\Delta G$ for this model (Eq.~\eqref{eqn:Gibbs-jensen}). For $P<2.0$ the agreement for all three curves with the analytical results is very good, while for high pressure the convergence is somewhat slower due to irreversible effects, which are more relevant for short switching times. We observed the convergence of our results to the analytical values in the case of the switching time $t_{sim}=10^5$ steps, which confirms the effectiveness of the method.
\subsection{Solid-solid phase transition in 2d a system}
As another example that shows the effectualness of the methodology we present a study that demonstrates the ability of the NPT Langevin dynamics with anisotropic volume fluctuations to produce a solid-solid phase transition. We chose a two-dimensional system with particles interacting through a potential with two minima,\cite{sspt} $U(r)$,  which is obtained by adding a Gaussian well to a smooth Lennard-Jones potential:\cite{smooth}
\begin{equation}
U(r)=4\epsilon\big[(\frac{\sigma}{r})^{12}-(\frac{\sigma}{r})^{6}+c_{2}(\frac{r}{\sigma})^{2}+c_{0}\big]-\lambda\epsilon e^{-\big(w\frac{r-r_{0}}{\sigma}\big)^{n}}.
\end{equation}
\begin{figure}[tbp]
    \centering
            \includegraphics[scale=0.40,bb=25 71 576 523,clip]{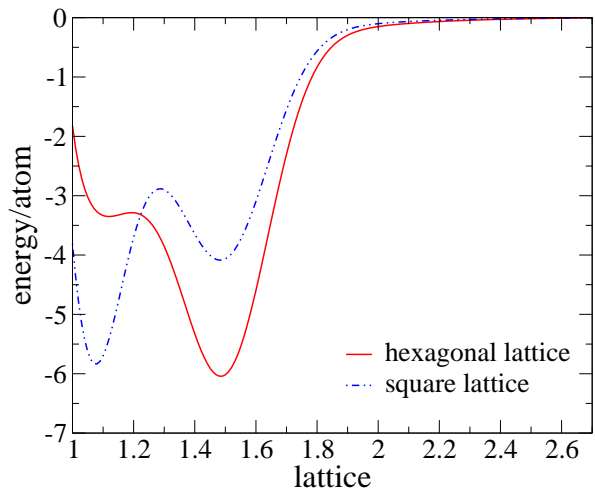}       
    \caption{\label{fig:potential} Energy per atom as function of lattice parameter.}
\end{figure}

In the calculations we use $\epsilon=1$, $\sigma=1$, $\lambda=1.7$, $w=5.0$, $r_{0}=1.5$, and $n=2$. The parameters $c_{0}$ and $c_{2}$ were chosen in such a way that the potential and the force become continuously zero at the truncation radius $r_{c}$. This kind of potential exhibits two stable crystalline structures: hexagonal and square lattices. Fig.~\ref{fig:potential} shows that the hexagonal lattice is more stable at low pressure than the square lattice. Therefore, a phase transition between the hexagonal and the square crystals is expected to occur as the system is compressed.

Initially, we perform a NPT(flex) simulation (Eqs.~\eqref{eqn:langevin-npt-flex}), beginning with the hexagonal structure in simulation cell containing $N=896$ particles.

The simulations were carried out using the following parameters, $T=0.462$, $P=0.0$, $\Delta t=0.01$, $\tau_{p}=0.5$, $\tau_{b}=5.0$, $r_{c}=3.0$ and the mass barostat is chosen according to the recommended formula:\cite{martyna} 
\begin{equation}
W_{g}=(dN+d)k_{B}T\tau^{2}_{b}/d.
\end{equation}

\begin{figure}[bp]
    \centering    
        \includegraphics[scale=0.28,bb=43 145 895 809,clip]{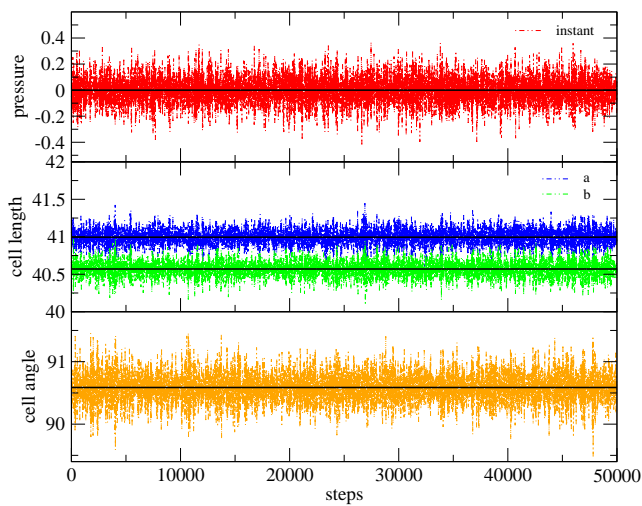}      
    \caption{\label{fig:thermo} (\textbf{Upper}) The instantaneous pressure vs. time. (\textbf{Center}) The instantaneous cell lengths (a, b) vs. time. (\textbf{Lower}) The instantaneous cell angles $\alpha$ vs. time. The black line are the average values.}
\end{figure}

\begin{figure}[tbp]
    \centering
        \includegraphics[scale=0.40,bb=29 133 576 522,clip]{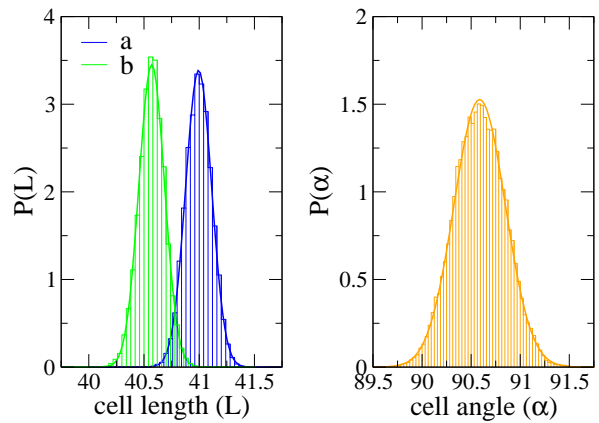}      
    \caption{\label{fig:distribution} (\textbf{Left}) The distribution of cell lengths a and b. (\textbf{Right}) The distribution of cell angle $\alpha$.}
\end{figure}

The system is equilibrated for $10^{5}$ steps and the production simulations used over $5\times10^{5}$ steps. The NPT(flex) simulation provides information about the fluctuations of the cell lengths and the cell angles individually. From the Figs.~\ref{fig:thermo} and ~\ref{fig:distribution}, we can see the fluctuations of cell lengths and angle, respectively. The edges a and b converge to 40.99 and 40.57, respectively. 

Each one of the edge lengths exhibit small fluctuations of less than 0.3. The cell angle $\alpha$ remains close to $90.59$ degrees with fluctuations of $\pm 0.75$ degrees. One can see that NPT(flex) simulation allows the fluctuation of cell lengths and angles, in a quasi-orthorhombic box shape.
Finally, we perform the compression-decompression  cycle of the system at constant  temperature $T=0.462$, beginning with the hexagonal structure at $P=0.0$ up to $P=1.8$, during a simulation time of $5\times 10^{6}$ steps; the same time interval is used to decompress the system.

\begin{figure}[bp]
    \centering
        \includegraphics[scale=0.40,bb=21 72 581 523,clip]{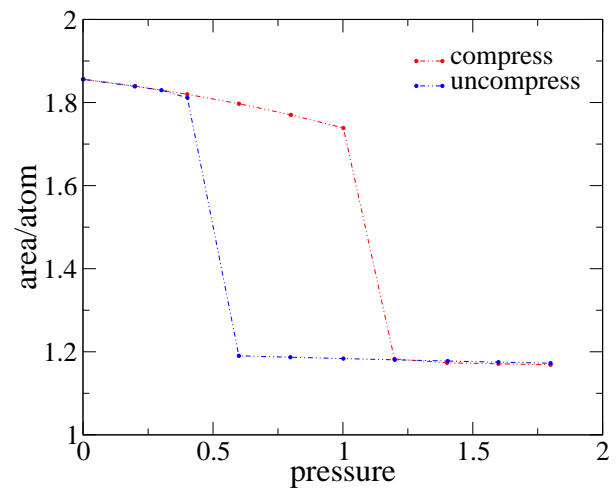}      
    \caption{\label{fig:loop} The atomic area as function of the external pressure applied at the system. The red(blue) line correspond to compress(uncompress) process.}
\end{figure}

\begin{figure}[tbp]
    \centering
    \begin{minipage}[b]{0.3\linewidth}
        \includegraphics[scale=0.22]{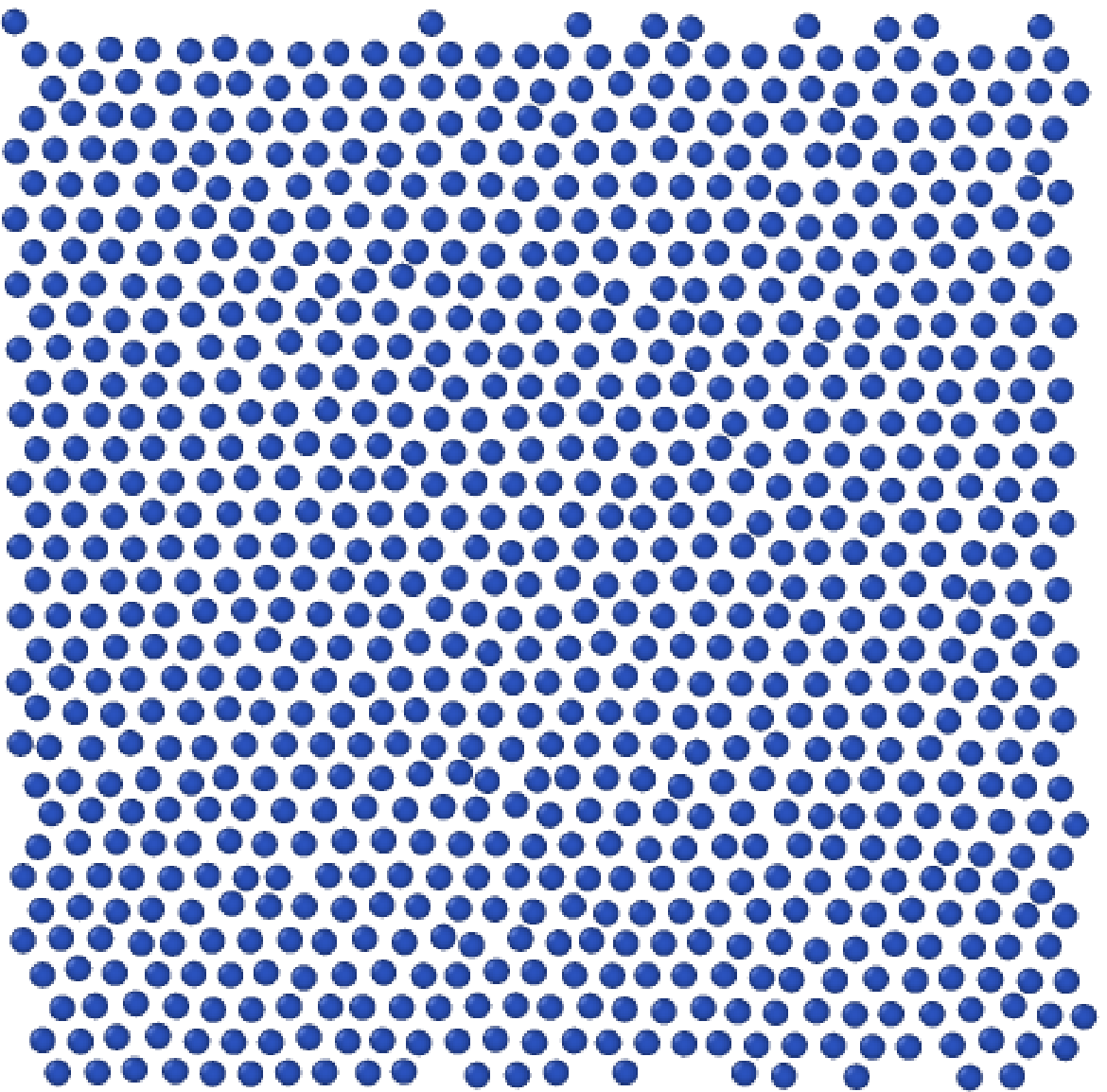}      
    \end{minipage}
    \begin{minipage}[b]{0.3\linewidth}
        \includegraphics[scale=0.24]{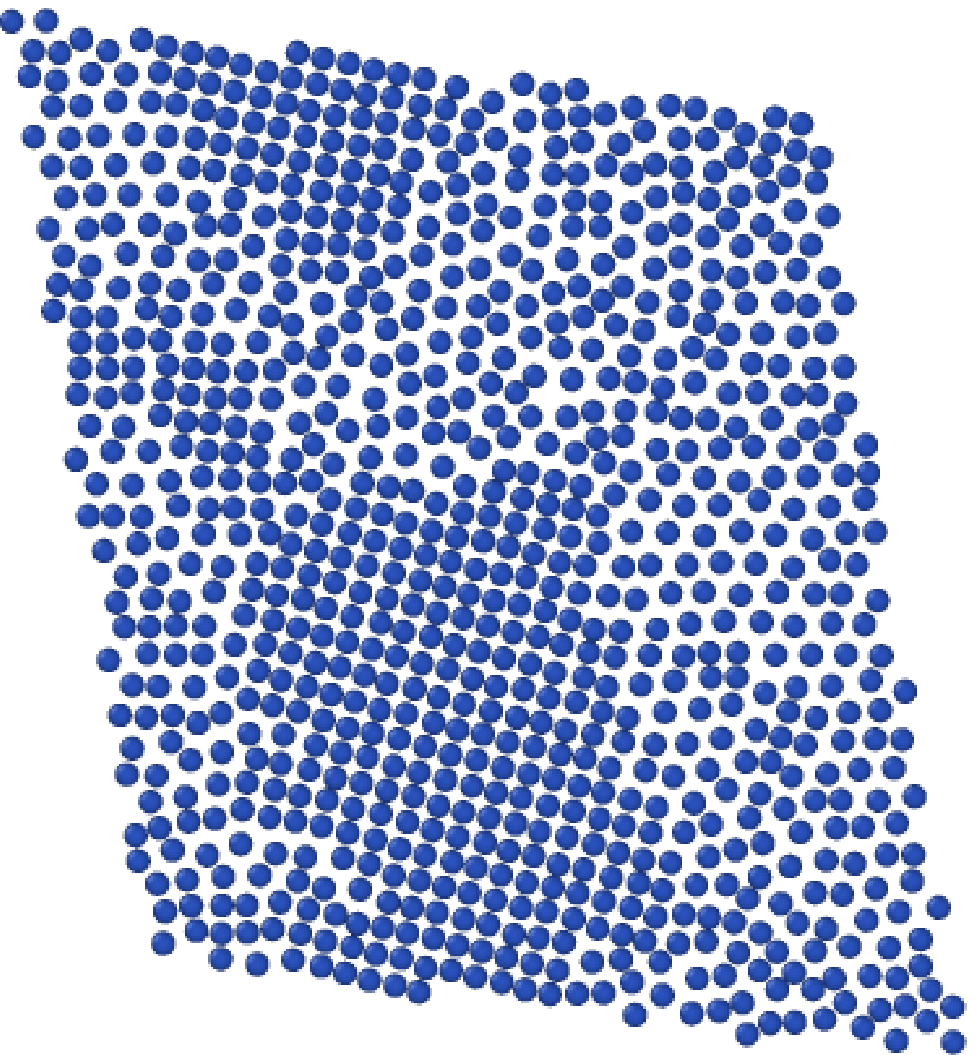}
    \end{minipage}
    \begin{minipage}[b]{0.3\linewidth}
        \includegraphics[scale=0.26]{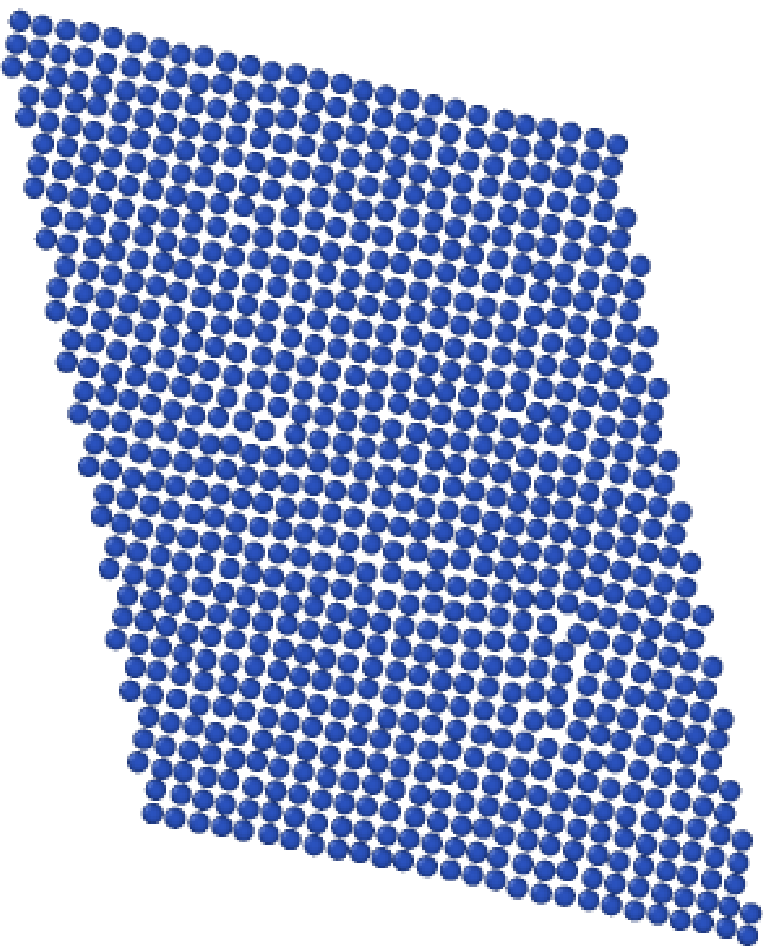}
    \end{minipage}
    \caption{(\textbf{Left}) Atomic configuration at low pressure.(\textbf{Center}) Atomic configuration at the phase transition. (\textbf{Rigth}) Atomic configuration at high pressure.}
    \label{fig:configurations}
\end{figure}

In Fig.~\ref{fig:loop}, we show the atomic area (area per
particle) as a function of the external pressure for a cycle of compression and decompression of the system. One can see the hysteresis loop, which is characteristic of a first-order phase transition. The atomic configurations showed in Fig.~\ref{fig:configurations} evidence the phase transition occurring between the hexagonal lattice and the square lattices, as expected to happen for the type of potential energy we used in the simulations (Fig.~\ref{fig:potential}).
The hysteresis loop in this case defines the region or interval of pressure where the phase transition takes place. However, if one is
interested in determining the transition pressure between the hexagonal and the square lattices, it is necessary to determine the crossing of the Gibbs free energy curves for each one of the phases.

In order to calculate the transition pressure, we used the methodology we have previously discussed to determine the Gibbs free energy in a wide interval of pressure at constant temperature.

The procedure to achieve this goal can be summarized as follows. Firstly, it is necessary to determine the reference Gibbs free energy for each crystalline phase. This is accomplished by the AS method using the Einstein crystal (a collection of harmonic oscillators) as a reference system whose free energy is known analytically.\cite{free,AS2,AS4} Since the Einstein crystal does not have cohesive energy, the AS simulations should be performed in the NVT ensemble.

\begin{figure}[bp]
	\centering	
	\includegraphics[scale=0.40,bb=7 72 591 515,clip]{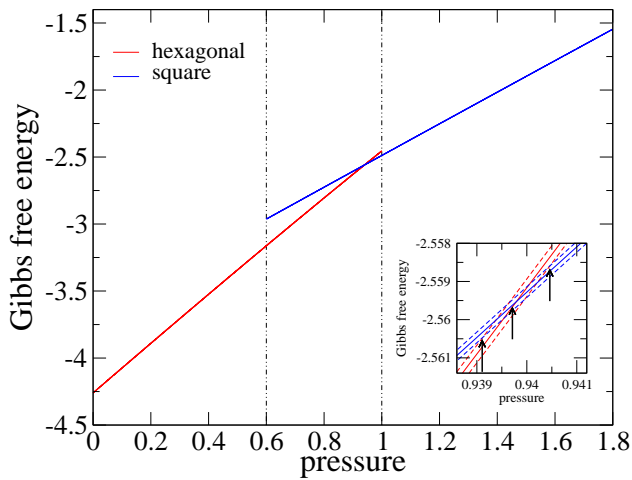}      
	\caption{\label{fig:gibbs-2d} AS-MD results for the Gibbs fee energy per particle as a function of pressure at the temperature of $T=0.462$. The intersection of the curves indicates the transition pressure at $P_{f}=0.939$, the dotted lines define a metastable region limits. (\textbf{Inset}) Gibbs free energy per particle around the transition pressure, for each phase we have three free energy curves: $\langle G \rangle$, $\langle G \rangle+\Delta G$ and $\langle G \rangle-\Delta G$, where $\Delta G$ is the total error calculated in the initial state point by the AS method. The arrows indicate the upper and lower limits of the transition pressure.}
\end{figure}
Initially, one has to perform NPT simulations at pressure values at which each one of the phases are stable. From these simulations one can determine the equilibrium volume (in our case equilibrium area) for each value of pressure. These equilibrium volumes are used in the AS simulations in the NVT ensemble. After determining the reference Gibbs free energy for each phase, one can use the AS method to compute the Gibbs free energy of each phase in a wide pressure interval. At last, the transition pressure is determined by the crossing of the two Gibbs free energy curves as shown in Fig.~\ref{fig:gibbs-2d}.

The phase transition occurs at the pressure $P=0.939$ at the constant temperature of $T=0.462$, in agreement with the metastable region indicated by the hysteresis loop previously determined. These results show that the correct sampling of the isothermal-isobaric ensemble, which is necessary to ensure the correct calculation of the Gibbs free energy, is achieved by the methodology proposed in this work. Furthermore, we propose a variant of the AS method that permits to determine efficiently the Gibbs free energy in a wide pressure interval at constant temperature using only a single simulation.
\section{Conclusions}\label{sec:Conclusions}
The sampling of the isobaric-isothermal ensemble has been re-examined by using the Langevin dynamics coupled to a barostat, which is described by non-Hamiltonian equations of motion. In the context of developing this new approach, we have extended the derivation of the Fokker-Planck equation for non-Hamiltonian systems, in order to describe the evolution of the distribution function in the extended phase-space. The numerical integration of this kind of equations of motion is deduced employing the Trotter scheme commonly used in the derivation of multiple-time-step integrators.

An analysis of the phase-space in these simulations has been presented, which shows that the correct probability distribution for the NPT ensemble with isotropic volume fluctuations is generated, by showing an excellent agreement with the results obtained for analytical models. We also performed simulations in the NPT ensemble with fully flexible cell, where it was possible to  observe a solid-solid phase transition, showing that the method is more suitable to simulate crystalline solids. 

The new methodology was used to determine the Gibbs free energy in a wide interval of pressure, which was shown to be in agreement with analytical results for the Jensen's model.\cite{jensen}

The methodology was applied to accurately and efficiently investigate the transition pressure of the hexagonal-to-square lattices phase transition, where the beginning and end states are chosen carefully in the AS simulations to reduce the error in the free energy calculation. 

\begin{acknowledgments}
We gratefully acknowledge support from the Brazilian agencies CNPq, CAPES, and FAPESP under Grants \#2010/16970-0 and \#2016/23891-6. The calculations were performed at CCJDR-IFGW-UNICAMP and at CENAPAD-SP in Brazil.
\end{acknowledgments}

\appendix

\section{Fokker-Planck equation for the NPT(iso) ensemble}\label{app:A}
This appendix contains the derivation of the Fokker-Planck equation that corresponds to a given NPT Langevin dynamics with isotropic volume fluctuations (the development is based on the book by R. Zwanzig\cite{Zwanzig} chapter 2). Let us start with a quite general Langevin equation for several variables $\{x_{1},x_{2},\ldots\}$, denoted for convenience by $\textbf{x}$, the equations of motion in this case are
\begin{equation}
\dot{x}_{i}=v_{i}(x_{i})+F_{i}(t),
\end{equation}
or, in an abbreviated form,
\begin{equation}
\dot{\textbf{x}}=\textbf{v}(\textbf{x})+\textbf{F}(t),
\end{equation}
where $\textbf{v}(\textbf{x})$ is some given function of the variables $\textbf{x}$. The noise $\textbf{F}(t)$ is Gaussian, with zero mean and the delta correlated second moment matrix,
\begin{equation}
\langle\textbf{F}(t)\textbf{F}(t')\rangle=2\textbf{B}\delta(t-t').
\end{equation}
and $\textbf{B}$ is a positive-semidefinite matrix denoting the thermal damping coefficient matrix. The associated Fokker-Planck equation for the averaged-noise distribution is:\cite{Zwanzig}
\begin{equation}
\frac{\partial}{\partial t}\rho(\textbf{x},t)=-\frac{\partial}{\partial\textbf{x}}\cdot\textbf{v}(\textbf{x})\rho(\textbf{x},t)+\frac{\partial}{\partial\textbf{x}}\cdot\textbf{B}\cdot\frac{\partial}{\partial\textbf{x}}\rho(\textbf{x},t),
\label{eqn:fkz}
\end{equation}
whose probability distribution should satisfy the condition
\begin{equation}
\int d\textbf{x}\rho(\textbf{x},t)=1\quad\forall t.
\end{equation}

In order to derive the Fokker-Planck equation that describe the evolution of the distribution function $\rho(\textbf{x},t)$, we can recognize the following quantities from the equations of motion Eq.~\eqref{eqn:langevin-npt-iso}, as the phase-space vector and the transpose of the differential operator as
\begin{equation}
\textbf{x}=\begin{pmatrix}
\textbf{r}_{1} \\
\vdots \\
\textbf{r}_{N} \\
\textbf{p}_{1} \\
\vdots \\
\textbf{p}_{N} \\
V \\
p_{\epsilon}
\end{pmatrix},
\qquad
\bigg[\frac{\partial}{\partial \textbf{x}}\bigg]^{T}=\begin{pmatrix}
\frac{\partial}{\partial {\textbf{r}}_{1}} \\
\vdots \\
\frac{\partial}{\partial {\textbf{r}}_{N}} \\
\frac{\partial}{\partial {\textbf{p}}_{1}} \\
\vdots \\
\frac{\partial}{\partial {\textbf{p}}_{N}}\\
\frac{\partial}{\partial V} \\
\frac{\partial}{\partial p_{\epsilon}}
\end{pmatrix},
\label{eqn:app-x-iso}
\end{equation}
the velocity of extended phase-space is
\begin{equation}
\textbf{v}(\textbf{x})=\begin{pmatrix}
\frac{\textbf{p}_{1}}{m_{1}}+\frac{p_{\epsilon}}{W}\textbf{r}_{1} \\
\vdots \\
\frac{\textbf{p}_{N}}{m_{N}}+\frac{p_{\epsilon}}{W}\textbf{r}_{N} \\
\textbf{f}_{1}-\big(1+\frac{d}{N_{f}}\big)\frac{p_{\epsilon}}{W}\textbf{p}_{1}-\gamma\textbf{p}_{1} \\
\vdots \\
\textbf{f}_{N}-\big(1+\frac{d}{N_{f}}\big)\frac{p_{\epsilon}}{W}\textbf{p}_{N}-\gamma\textbf{p}_{N} \\
\frac{dV}{W}p_{\epsilon} \\
G_{\epsilon}
\end{pmatrix}.
\end{equation}

For convenience we can separate the generalized velocity in a deterministic and a drift components:
\begin{equation}
\textbf{v}(\textbf{x})=\textbf{v}_{det}(\textbf{x})+\textbf{v}_{drift}(\textbf{x}),
\label{eqn:app-v-iso}
\end{equation}
where,
\begin{equation}
\textbf{v}_{det}(\textbf{x})=\begin{pmatrix}
\frac{\textbf{p}_{1}}{m_{1}}+\frac{p_{\epsilon}}{W}\textbf{r}_{1} \\
\vdots \\
\frac{\textbf{p}_{N}}{m_{N}}+\frac{p_{\epsilon}}{W}\textbf{r}_{N} \\
\textbf{f}_{1}-\big(1+\frac{d}{N_{f}}\big)\frac{p_{\epsilon}}{W}\textbf{p}_{1} \\
\vdots \\
\textbf{f}_{N}-\big(1+\frac{d}{N_{f}}\big)\frac{p_{\epsilon}}{W}\textbf{p}_{N} \\
\frac{dV}{W}p_{\epsilon} \\
G_{\epsilon}
\end{pmatrix},
\textbf{v}_{drift}(\textbf{x})=\begin{pmatrix}
0 \\
\vdots \\
0 \\
-\gamma \textbf{p}_{1} \\
\vdots \\
-\gamma \textbf{p}_{N} \\
0 \\
-\gamma_{b}p_{\epsilon} \\
\end{pmatrix}
\end{equation}
and the noise vector is
\begin{equation}
\textbf{F}(t)=\begin{pmatrix}
0 \\
\vdots \\
0 \\
\textbf{R}_{1}(t) \\
\vdots \\
\textbf{R}_{N}(t) \\
0 \\
R_{b}(t)
\end{pmatrix},
\end{equation}

The first term of the right-hand side of Eq.~\eqref{eqn:fkz} can be split as:
\begin{align}
&-\frac{\partial}{\partial\textbf{x}}\cdot\textbf{v}(\textbf{x})\rho(\textbf{x},t)=-\bigg(\underbrace{\frac{\partial}{\partial\textbf{x}}\cdot\textbf{v}_{det}(\textbf{x})}_{k(\textbf{x})=0}\bigg)\rho(\textbf{x},t)
\nonumber \\
&+\underbrace{\textbf{v}_{det}(\textbf{x)}\cdot\frac{\partial}{\partial\textbf{x}}}_{\mathcal{L}}\rho(\textbf{x},t)-\frac{\partial}{\partial\textbf{x}}\cdot\textbf{v}_{drift}(\textbf{x})\rho(\textbf{x},t).
\label{eqn:v-fk-iso}
\end{align}

It is easy to show that the compressibility of the deterministic term in the equations of motion is zero and $\mathcal{L}$ is the corresponding Liouville operator. Using Eqs.~\eqref{eqn:app-x-iso} the Liouville operator $\mathcal{L}$ operator can be written as
\begin{align}
&\textbf{v}_{det}(\textbf{x)}\cdot\frac{\partial}{\partial\textbf{x}}=\sum_{i=1}^{N}\bigg[\bigg(\frac{\textbf{p}_{i}}{m_{i}}+\frac{p_{\epsilon}}{W}\textbf{r}_{i}\bigg)\cdot\frac{\partial}{\partial \textbf{r}_{i}} \\
&+\bigg(\textbf{f}_{i}-\Big(1+\frac{d}{N_{f}}\Big)\frac{p_{\epsilon}}{W}\textbf{p}_{i}\bigg)\cdot\frac{\partial}{\partial \textbf{p}_{i}}\bigg] +\frac{dV}{W}p_{\epsilon}\frac{\partial}{\partial V}+G_{\epsilon}\frac{\partial}{\partial p_{\epsilon}}.\nonumber 
\label{eqn:v-det-iso}
\end{align}

Finally the operator associated to drift and noise is:
\begin{align}
&-\frac{\partial}{\partial\textbf{x}}\cdot\textbf{v}_{drift}(\textbf{x})+\frac{\partial}{\partial\textbf{x}}\cdot\textbf{B}\cdot\frac{\partial}{\partial\textbf{x}}=\gamma\sum_{i=1}^{N}\frac{\partial}{\partial \textbf{p}_{i}}\cdot\bigg(\textbf{p}_{i} \nonumber \\
&+m_{i}k_{B}T\frac{\partial}{\partial \textbf{p}_{i}}\bigg)+\gamma_{b}\frac{\partial}{\partial p_{\epsilon}}\big(p_{\epsilon}+Wk_{B}T\frac{\partial}{\partial p_{\epsilon}}\big). 
\end{align}

Combining these last results the Fokker-Planck equation for the NPT ensemble with isotropic volume fluctuations is obtained,  
\begin{equation}
\frac{\partial }{\partial t}\rho(\textbf{x},t)=-\mathcal{L}_{NPT(iso)}\rho(\textbf{x},t),
\end{equation}
where, 
\begin{align}
&\mathcal{L}_{NPT(iso)}=\sum_{i=1}^{N}\bigg[\bigg(\textbf{f}_{i}-\Big(1+\frac{d}{N_{f}}\Big)\frac{p_{\epsilon}}{W}\textbf{p}_{i}\bigg)\cdot\frac{\partial}{\partial \textbf{p}_{i}} \nonumber \\
&+ \bigg(\frac{\textbf{p}_{i}}{m_{i}}+\frac{p_{\epsilon}}{W}\textbf{r}_{i}\bigg)\cdot\frac{\partial}{\partial \textbf{r}_{i}}\bigg]+\frac{dV}{W}p_{\epsilon}\frac{\partial}{\partial V} +G_{\epsilon}\frac{\partial}{\partial p_{\epsilon}} \nonumber \\
&-\gamma\sum_{i=1}^{N}\frac{\partial}{\partial \textbf{p}_{i}}\cdot\bigg(\textbf{p}_{i}+m_{i}k_{B}T\frac{\partial}{\partial \textbf{p}_{i}}\bigg) \nonumber \\
&-\gamma_{b}\frac{\partial}{\partial p_{\epsilon}}\big(p_{\epsilon}+Wk_{B}T\frac{\partial}{\partial p_{\epsilon}}\big).
\label{eqn:fokker-planck-npt-iso}
\end{align}
\section{Fokker-Planck equation for the NPT(flex) ensemble}\label{app:B}
For the fully flexible case, from the equations of motion (Eq.~\eqref{eqn:langevin-npt-flex}), we can identify the following terms: the phase-space vector and the transpose of the differential operator as
\begin{equation}
\textbf{x}=\begin{pmatrix}
\textbf{r}_{1} \\
\vdots \\
\textbf{r}_{N} \\
\textbf{p}_{1} \\
\vdots \\
\textbf{p}_{N} \\
\textbf{h} \\
\textbf{p}_{g} \\
\end{pmatrix},
\qquad
\bigg[\frac{\partial}{\partial \textbf{x}}\bigg]^{T}=\begin{pmatrix}
\frac{\partial}{\partial {\textbf{r}}_{1}} \\
\vdots \\
\frac{\partial}{\partial {\textbf{r}}_{N}} \\
\frac{\partial}{\partial {\textbf{p}}_{1}} \\
\vdots \\
\frac{\partial}{\partial {\textbf{p}}_{N}}\\
\frac{\partial}{\partial \textbf{h}} \\
\frac{\partial}{\partial \textbf{p}_{g}}
\end{pmatrix},
\label{eqn:app-x-flex}
\end{equation}
we also can separate the velocity of extended phase-space in deterministic and drift components:
\begin{equation}
\textbf{v}_{det}(\textbf{x})=\begin{pmatrix}
\frac{\textbf{p}_{1}}{m_{1}}+\frac{\textbf{p}_{g}}{W_{g}}\textbf{r}_{1} \\
\vdots\\
\frac{\textbf{p}_{N}}{m_{N}}+\frac{\textbf{p}_{g}}{W_{g}}\textbf{r}_{N} \\
\textbf{f}_{1}-\frac{\textbf{p}_{g}}{W_{g}}\textbf{p}_{1}-\frac{1}{N_{f}}\frac{\mathrm{tr}(\textbf{p}_{g})}{W_{g}}\textbf{p}_{1} \\
\vdots \\
\textbf{f}_{N}-\frac{\textbf{p}_{g}}{W_{g}}\textbf{p}_{N}-\frac{1}{N_{f}}\frac{\mathrm{tr}(\textbf{p}_{g})}{W_{g}}\textbf{p}_{N} \\
\frac{\textbf{p}_{g}\textbf{h}}{W_{g}} \\
\textbf{G}_{g}
\end{pmatrix},
\end{equation}
\begin{equation}
\textbf{v}_{drift}(\textbf{x})=\begin{pmatrix}
0 \\
\vdots \\
0 \\
-\gamma \textbf{p}_{1}\\
\vdots \\
-\gamma \textbf{p}_{N}\\
0 \\
-\gamma_{b}\textbf{p}_{g}
\end{pmatrix},
\quad
\textbf{F}(t)=\begin{pmatrix}
0 \\
\vdots \\
0 \\
\textbf{R}_{1} \\
\vdots \\
\textbf{R}_{N}\\
0 \\
\textbf{R}_{g}
\end{pmatrix},
\end{equation}
where $\textbf{F}(t)$ is the noise vector.

When the deterministic component has non-null compressibility, the extension of the formulation of the Fokker-Planck equation is not straightforward. To this end, similar ideas to those proposed by Tuckerman\cite{BookTuckerman} can be used, by introducing for convenience a phase-space metric term $[\mathrm{det}(\textbf{h})]^{1-d}$, in such a way that in the steady state regime the solution of the Fokker-Planck equation reproduces the probability distribution function of the isothermal-isobaric ensemble in the fully flexibly cell case. 

The probability condition is 
\begin{equation}
\int d\textbf{x}[\mathrm{det}(\textbf{h})]^{1-d}\rho(\textbf{x},t)=1\quad\forall t,
\end{equation}
the Fokker-Planck equations take the form:
\begin{align}
&\frac{\partial }{\partial t}\bigg([\mathrm{det}(\textbf{h})]^{1-d}\rho(\textbf{x},t)\bigg)=-\frac{\partial}{\partial\textbf{x}}\cdot\bigg(\textbf{v}(\textbf{x})[\mathrm{det}(\textbf{h})]^{1-d}\rho(\textbf{x},t)\bigg)\nonumber \\
&+\frac{\partial}{\partial\textbf{x}}\cdot\textbf{B}\cdot\frac{\partial}{\partial\textbf{x}}\bigg([\mathrm{det}(\textbf{h})]^{1-d}\rho(\textbf{x},t)\bigg).
\label{eqn:fkz-flex}
\end{align}
%
The left hand side of Eq.~\eqref{eqn:fkz-flex} can be written as:
\begin{equation}
\frac{\partial }{\partial t}\bigg([\mathrm{det}(\textbf{h})]^{1-d}\rho(\textbf{x},t)\bigg)=[\mathrm{det}(\textbf{h})]^{1-d}\frac{\partial }{\partial t}\rho(\textbf{x},t),
\end{equation}
%
and the first right hand side term of Eq.~\eqref{eqn:fkz-flex} is
\begin{align}
&-\frac{\partial}{\partial\textbf{x}}\cdot\bigg(\textbf{v}(\textbf{x})[\mathrm{det}(\textbf{h})]^{1-d}\rho(\textbf{x},t)\bigg)=-\frac{\partial}{\partial\textbf{x}}\cdot\bigg(\textbf{v}_{det}(\textbf{x}) \nonumber \\
&\times[\mathrm{det}(\textbf{h})]^{1-d}\rho(\textbf{x},t)\bigg) \nonumber \\
&-\frac{\partial}{\partial\textbf{x}}\cdot\bigg(\textbf{v}_{drift}(\textbf{x})[\mathrm{det}(\textbf{h})]^{1-d}\rho(\textbf{x},t)\bigg).
\label{eqn:v-flex}
\end{align}
%
Expanding the first term on the right hand side of Eq.~\eqref{eqn:v-flex} we obtain
\begin{align}
&\frac{\partial}{\partial\textbf{x}}\cdot\bigg(\textbf{v}_{det}(\textbf{x})[\mathrm{det}(\textbf{h})]^{1-d}\rho(\textbf{x},t)\bigg)=\underbrace{\frac{\partial}{\partial\textbf{x}}\cdot\textbf{v}_{det}(\textbf{x})}_{k(\textbf{x})=-(1-d)\frac{\mathrm{tr}(\textbf{p}_{g})}{W_{g}}\neq 0}\nonumber \\
&\times[\mathrm{det}(\textbf{h})]^{1-d}\rho(\textbf{x},t)+\underbrace{\textbf{v}_{det}(\textbf{x)}\cdot\bigg(\frac{\partial}{\partial\textbf{x}}[\mathrm{det}(\textbf{h})]^{1-d}\bigg)}_{(1-d)\frac{\mathrm{tr}(\textbf{p}_{g})}{W_{g}}[\mathrm{det}(\textbf{h})]^{1-d}}\rho(\textbf{x},t)\nonumber \\
&-[\mathrm{det}(\textbf{h})]^{1-d}\textbf{v}_{det}(\textbf{x)}\cdot\frac{\partial}{\partial\textbf{x}}\rho(\textbf{x},t),
\label{eqn:v-det}
\end{align}
%
which results in
\begin{align}
&-\frac{\partial}{\partial\textbf{x}}\cdot\bigg(\textbf{v}_{det}(\textbf{x})[\mathrm{det}(\textbf{h})]^{1-d}\rho(\textbf{x},t)\bigg)=-[\mathrm{det}(\textbf{h})]^{1-d} \nonumber \\
&\times\textbf{v}_{det}(\textbf{x)}\cdot\frac{\partial}{\partial\textbf{x}}\rho(\textbf{x},t),
\end{align}
since the first and second terms on the right hand side of Eq.~\eqref{eqn:v-det} cancel each other.

%
By adding the second term on the right hand side of Eq.~\eqref{eqn:v-flex} to noise terms we have:
\begin{align}
&-\frac{\partial}{\partial\textbf{x}}\cdot\bigg(\textbf{v}_{drift}(\textbf{x})[\mathrm{det}(\textbf{h})]^{1-d}\rho(\textbf{x},t)\bigg) \nonumber \\
&+\frac{\partial}{\partial\textbf{x}}\cdot\textbf{B}\cdot\frac{\partial}{\partial\textbf{x}}\bigg([\mathrm{det}(\textbf{h})]^{1-d}\rho(\textbf{x},t)\bigg)= \nonumber \\
&+\gamma\sum_{i=1}^{N}\frac{\partial}{\partial \textbf{p}_{i}}\cdot\bigg(\textbf{p}_{i}+m_{i}k_{B}T\frac{\partial}{\partial \textbf{p}_{i}}[\mathrm{det}(\textbf{h})]^{1-d}\rho(\textbf{x},t)\bigg)\nonumber \\
&+\gamma_{b}\frac{\partial}{\partial \textbf{p}_{g}}\cdot\bigg(\textbf{p}_{g}+W_{g}k_{B}T\frac{\partial}{\partial \textbf{p}_{g}}[\mathrm{det}(\textbf{h})]^{1-d}\rho(\textbf{x},t)\bigg). 
\end{align}
%



Using these results in Eq.~\eqref{eqn:fkz-flex} and canceling out the phase-space metric factor present in all terms of the equation ($[\mathrm{det}(\textbf{h})]^{1-d}\neq0$), one can obtain the Fokker-Planck equation to describe the NPT Langevin dynamics with anisotropic volume fluctuations for the unweighted distribution function

\begin{equation}
\frac{\partial}{\partial t}\rho(\textbf{x},t)=-\mathcal{L}_{NPT(flex)}\rho(\textbf{x},t),
\end{equation}
where,
\begin{align}
&\mathcal{L}_{NPT(flex)}=\sum_{i=1}^{N}\bigg[\bigg(\textbf{f}_{i}-\frac{\textbf{p}_{g}}{W_{g}}\textbf{p}_{i}-\frac{1}{N_{f}}\frac{\mathrm{Tr}(\textbf{p}_{g})}{W_{g}}\textbf{p}_{i}\bigg)\cdot\frac{\partial}{\partial \textbf{p}_{i}} \nonumber \\
&+\bigg(\frac{\textbf{p}_{i}}{m_{i}}+\frac{\textbf{p}_{g}}{W_{g}}\textbf{r}_{i}\bigg)\cdot\frac{\partial}{\partial \textbf{r}_{i}}\bigg] +\bigg(\frac{\textbf{p}_{g}\textbf{h}}{W_{g}}\bigg)\cdot\frac{\partial}{\partial\textbf{h}}+\textbf{G}_{g}\cdot\frac{\partial}{\partial\textbf{p}_{g}} \nonumber \\ 
&-\gamma\sum_{i=1}^{N}\frac{\partial}{\partial \textbf{p}_{i}}\cdot\bigg(\textbf{p}_{i}+m_{i}k_{B}T\frac{\partial}{\partial \textbf{p}_{i}}\bigg) \nonumber \\
&-\gamma_{b}\frac{\partial}{\partial \textbf{p}_{g}}\cdot\bigg(\textbf{p}_{g}+W_{g}k_{B}T\frac{\partial}{\partial \textbf{p}_{g}}\bigg).
\label{eqn:fokker-planck-npt-flex}
\end{align}
\bibliographystyle{apsrev4-1}
\end{document}